\documentclass{iopart}

\usepackage{iopams}%
\usepackage{graphicx}%

\begin{document}

\title[Dust ion acoustic solitary structures]{Dust ion acoustic solitary structures in nonthermal dusty plasma}

\author{Anup Bandyopadhyay$^{1}$, Animesh Das$^{1}$ and K P Das$^{2}$}%
\address{$^{1}$ Department of Mathematics, Jadavpur
University, Kolkata - 700 032, India.
}
\address{$^{2}$ Department of Applied Mathematics,
University of Calcutta, 92 - Acharya Prafulla Chandra Road, Kolkata -
700 009, India}
\ead{ab\_ju\_math@yahoo.co.in}
\begin{abstract}
Dust ion acoustic solitary structures have been investigated in an unmagnetized nonthermal plasma consisting of negatively charged dust grains, adiabatic positive ions and nonthermal electrons. For isothermal electrons, the present plasma system does not support any double layer solution, whereas for nonthermal electrons, negative potential double layer starts to occur whenever the nonthermal parameter exceeds a critical value. However this double layer solution is unable to restrict the occurrence of all negative potential solitary waves of the present system. As a result, two different types of negative potential solitary waves have been observed, in which occurrence of first type of solitary wave is restricted by $M_{c}<M<M_{D}$ whereas the second type solitary wave exists for all $M>M_{D}$, where $M_{c}$ is the lower bound of Mach number $M$, i.e., solitary structures start to exist for $M>M_{c}$ and $M_{D}(>M_{c})$ is the Mach number corresponding to a negative potential double layer. A finite jump between the amplitudes of negative potential of solitary waves at $M=M_{D}-\epsilon_{1}$ and at $M=M_{D}+\epsilon_{2}$ has been observed, where $0<\epsilon_{1}<M_{D}-M_{c}$ and $\epsilon_{2}>0$. As double layer solution plays an important role for the present system, an analytical theory for the existence of double layer has been presented. A numerical scheme has also been provided to find the value of Mach number at which double layer solution exists and also the amplitude of that double layer. The solitary structures of both polarities can coexist whenever $\mu$ exceeds a critical value, where $\mu$ is the ratio of the unperturbed number density of electrons to that of ions. Although the occurrence of coexistence of solitary structures of both polarities is restricted by $M_{c}<M \leq M_{max}$, only negative potential solitary wave still exists for all $M>M_{max}$, where $M_{max}$ is the upper bound of $M$ for the existence of positive potential solitary waves only. Qualitatively different solution spaces, i.e., the compositional parameter spaces showing the nature of existing solitary structures of the energy integral have been found. These solution spaces are capable of producing new results and physical ideas for the formation of solitary structures whenever one can move the solution spaces through the family of curves parallel to the curve $M=M_{c}$.
\end{abstract}
\maketitle

\section{\label{sec:intro}Introduction}
Acoustic wave modes in dusty plasma have received a great deal of attention since the last decade  \cite{Rao90,Shukla92,Verheest92,Barkan95,Pieper96,Mamun96a,Mamun96b,Shukla01}. Depending on different time scales, there can exists two or more acoustic waves in a typical dusty plasma. Dust Acoustic (DA) and Dust Ion-Acoustic (DIA) waves are two such acoustic waves in a plasma containing electrons, ions, and charged dust grains.

Shukla and Silin \cite{Shukla92} were the first to show that due to the quasi neutrality condition $n_{e0}+n_{d0}Z_{d} = n_{i0}$ and the strong inequality $n_{e0} \ll n_{i0}$ ($n_{e0}$, $n_{i0}$, and $n_{d0}$ are, respectively, the number density of electrons, ions, and dust particles, where $Z_{d}$ is the number of electrons residing on the dust grain surface), a dusty plasma (with negatively charged static dust grains) supports low-frequency DIA waves with phase velocity much smaller (larger) than electron (ion) thermal velocity. In case of long wavelength limit the dispersion relation of DIA wave is similar to that of Ion-Acoustic (IA) wave for a plasma with $n_{e0} = n_{i0}$ and $T_{i} \ll T_{e}$, where $T_{i} (T_{e})$ is the average ion (electron) temperature. Due to the usual dusty plasma approximations ($n_{e0} \ll n_{i0}$ and $T_{i}\simeq T_{e}$), a dusty plasma cannot support the usual IA waves, but the DIA waves of Shukla and Silin \cite{Shukla92} can. Thus DIA waves are basically  IA waves, modified by the presence of heavy dust particulates. The theoretical prediction of Shukla and Silin \cite{Shukla92} was supported by a number of laboratory experiments \cite{Barkan96,Merlino98,Nakamura01}. The linear properties of DIA waves in dusty plasma are now well understood \cite{Shukla99,Verheest00,Shukla02}.

Dust Ion-Acoustic solitary waves (DIASWs) have been investigated by several authors. Bharuthram and Shukla \cite{Bharuthram92} studied the DIASWs in an unmagnetized dusty plasma consisting of isothermal electrons, cold ions, in both static and mobile dust particles. Employing reductive perturbation method, Mamun and Shukla \cite{Mamun02} investigated the cylindrical and spherical DIASWs in an unmagnetized dusty plasma consisting of inertial ions, isothermal electrons, and stationary dust particles. They \cite{Mamun02psc} have also investigated the condition for existence of positive and negative potential DIASWs. Verheest \textit{et al.} \cite{Verheest05} have shown that in the dust-modified ion acoustic regime, negative structures can also be generated, beside positive potential soliton if the polytropic index $\gamma_{e} \neq 1$ for electrons. The effect of ion-fluid temperature on DIASWs structures have been investigated by Sayed and Mamun \cite{Sayed08} in a dusty plasma containing adiabatic ion-fluid, Boltzmann electrons, and static dust particles.

In most of the earlier works, Maxwellian velocity distribution function for lighter species of particles has been used to study DIASWs and DIA double layers (DIADLs). However the dusty plasma with nonthermally/suprathermally  distributed electrons observed in a number of heliospheric environments \cite{Verheest00,Shukla02,Asbridge68,Feldman83,Lundin89,Futaana03}. Therefore, it is of considerable importance to study nonlinear wave structures in a dusty plasma in which lighter species (electrons) is nonthermally/suprathermally distributed. Berbri and Tribeche \cite{Berbri09} have investigated weakly nonlinear DIA shock waves in a dusty plasma with nonthermal electrons. Recently Baluku \textit{et al.} \cite{Baluku10a} have investigated DIASWs in an unmagnetized dusty plasma consisting of cold dust particles and kappa distributed electrons using both small and arbitrary amplitude techniques.

In the present investigation we have considered the problem of existence of DIASWs and DIADLs in a plasma consisting of negatively charged dust grains, adiabatic positive ions and nonthermal electrons. Three basic parameters of the present dusty plasma system are $\mu$, $\alpha$ and $ \beta_{1}$, which are respectively the ratio of unperturbed number density of nonthermal electrons to that of ions, the ratio of average temperature of ions to that of nonthermal electrons, a parameter associated with the nonthermal distribution of electrons. Nonthermal distribution of electrons becomes isothermal one if $\beta_{1}=0$.

The main aim of this paper is to investigated DIASWs and DIADLs thoroughly, giving special emphasis on the followings:\\
(\textbf{a}) To study the nonlinear properties of DIA waves in a dusty plasma with nonthermal electrons. (\textbf{b}) To find the exact bounds (lower and upper) of the Mach number $M$ for the existence of solitary wave solutions. (\textbf{c}) As double layer solution plays an important role to restrict the occurrence of at least one sequence of solitary waves of same polarity, we set up an analytical theory  to find the double layer solution of the energy integral, which help us to find the Mach number at which double layer occurs and also, to find the amplitude of that double layer solution. (\textbf{d}) On the basis of the analytical theory for the existence of solitary waves and double layers, the present plasma system has been analyzed numerically. Actually, analyzing the Sagdeev potential, we have found qualitatively different solution spaces or the compositional parameter spaces showing the nature of existing solitary structures of the energy integral. From these solution spaces, the main observations are the followings. (\textbf{d1}) For isothermal electrons, the present plasma system does not support any double layer solution in both cold and adiabatic cases. For nonthermal electrons, the present plasma system does not support any Positive Potential Double Layer (PPDL) solution, whereas Negative Potential Double Layers (NPDLs) start to occur whenever the nonthermal parameter exceeds a critical value. However this NPDL solution is unable to restrict the occurrence of all Negative Potential Solitary Waves (NPSWs) of the present system, i.e., NPDL solution is not the ultimate solution of the energy integral in the negative potential side. Actually, we have observed two different types of NPSWs, in which amplitude of first type of NPSW is restricted by the amplitude of NPDL whereas the amplitude of NPDL is unable to restrict the amplitude of the second type NPSWs. As a result, we have observed a finite jump in amplitudes between two different types of NPSWs separated by a NPDL. This fact has also been observed recently by Verheest\cite{Verheest10b} and Baluku \textit{et al.}\cite{Baluku10b} for ion-acoustic solitary wave with different plasma constituents. (\textbf{d2}) For any physically admissible values of the parameters of the system, specifically, for any value of $\mu$ and any value of $\beta_{1}$, NPSW exists for all $M>M_{c}$ except $M=M_{D}$, where $M_{c}$ is the lower bound of Mach number $M$, i.e., solitary structures start to exist for $M>M_{c}$ and $M_{D}(>M_{c})$ is the Mach number corresponding to a NPDL solution. However, if the parameter $\mu$ exceeds a critical value $\mu_{p}$, PPSWs exist for all $\mu_{p} \leq \mu \leq \mu_{T}( < 1)$ whenever the Mach number lies within the interval $M_{c} < M \leq M_{max}$, where $\mu_{T}( < 1)$ is a physically admissible upper bound of $\mu$ and $M_{max}$ is the upper bound of $M$ for the existence of PPSWs only, i.e., there does not exist any PPSW if $M>M_{max}$. Therefore, the coexistence of both PPSWs and NPSWs is possible for all $\mu_{p} \leq \mu \leq \mu_{T}$ whenever $M_{c} < M \leq M_{max}$, but NPSWs still exist for $M>M_{max}$. (\textbf{d3}) For nonthermal electron species, we have investigated the entire solution space of the energy integral with respect to the nonthermal parameter $\beta_{1}$ and we have found four qualitatively different solution spaces depending on the cut off values of $\mu$. Actually, here we are able to define three cut off values $\mu_{p}$, $\mu_{q}$ and $\mu_{r}$ of $\mu$ such that $0<\mu_{p}<\mu_{q}<\mu_{r}<1$ for any given value of $\alpha$, and consequently, we can partition the entire range of $\mu$ in the following four disjoint subintervals: $0<\mu<\mu_{p}$, $\mu_{p}\leq \mu <\mu_{q}$, $\mu_{q}\leq \mu <\mu_{r}$ and $\mu_{r}\leq \mu \leq \mu_{T}( <1)$. For these four disjoint subintervals of $\mu$, we have four different solution spaces of the energy integral with respect to nonthermal parameter $\beta_{1}$. These solution spaces can define all types of solitary structures of the present system. (\textbf{d4}) Finally, considering any solution space, one can get new results and physical ideas for the formation of solitary structures if he moves the solution space through the family of curves parallel to the curve $M=M_{c}$. If we move the solution space through the family of curves parallel to the curve $M=M_{c}$, it is simple to understand the mathematics as well as physics for the formation of double layer solution and it is also simple to understand the relation between solitons and double layer solution.

The present paper is organized as follows: Basic equations are given in \sref{sec:non eqn}. Derivation of energy integral along with Sagdeev potential is given in \sref{sec:Energy int}. Physical interpretation for the existence of solitary structures of the energy integral is given in \sref{sec:physics}. The lower and upper bounds of the Mach number for the existence of solitary structures are given in \sref{sec:mach number}. In  \sref{subsec:positive}, the analytical method to find the upper bound of the Mach number for the existence of PPSWs is given. In \sref{subsec:negative}, we find that the existence of NPDL solution may restrict occurrence of NPSWs having amplitude less than the amplitude of NPDL. In \sref{subsec:dl}, an analytical theory to find the double layer solution of the energy integral has been provided. In \sref{algo}, an algorithm has been provided to find the value of Mach number at which double layer solution exists and also the amplitude of that double layer. Following a logical sequence of numerical scheme based on the theoretical discussions as given in \sref{sec:physics} and \sref{sec:mach number}, the solution spaces have been constructed in \sref{sec:solution space}. Finally, we have concluded our findings in \sref{sec:conclusion}.

\section{\label{sec:non eqn}Basic equations}
The governing equations describing the nonlinear behavior of DIA waves, propagating along $x$-axis, in collisionless, unmagnetized dusty plasma consisting of negatively charged immobile dust grains are the following:
\begin{eqnarray}\label{continuity of ions}
n_{i,t}+ (n_{i} u_{i})_{x} = 0,
\end{eqnarray}
\begin{eqnarray}\label{motion of ions}
u_{i,t}+u_{i}u_{i,x} = -\phi_{x} - \alpha n_{i}^{-1}p_{i,x},
\end{eqnarray}
\begin{eqnarray}\label{pressure eq}
p_{i,t}+u_{i}p_{i,x}+ \gamma p_{i}u_{i,x} = 0,
\end{eqnarray}
\begin{eqnarray}\label{Poisson eq without mu}
\phi_{xx} = (n_{e0}/n_{i0}) n_{e}- n_{i} + (Z_{d} n_{d0})/n_{i0},
\end{eqnarray}
where the parameter $\alpha = T_{i}/T_{e}$.

Here we have used the notation $\psi_{i,~q}$ or $(\psi_{i})_{q}$ for $\partial \psi_{i}/\partial q$ and $n_{i}$, $n_{e}$, $u_{i}$, $p_{i}$, $\phi$, $x$ and $t$ are, respectively, the ion number density, electron number density, ion velocity, ion pressure, electrostatic potential, spatial variable and time, and they have been normalized by $n_{i0}$ (unperturbed ion number density), $n_{e0}$ (unperturbed electron number density), $c_{i}$($=\sqrt{(K_{B}T_{e})/m_{i}}$) (ion-acoustic speed), $ n_{i0}K_{B}T_{i}$, $K_{B}T_{e}/e$, $\lambda_{Dem}(=\sqrt{(K_{B}T_{e})/(4\pi n_{i0} e^{2})})$(Debye length), and $\omega_{pi}^{-1}(=\sqrt{m_{i}/(4\pi n_{i0} e^{2})})$ (ion plasma period). Here $\gamma (=3)$ is the adiabatic index, $K_{B}$ is the Boltzmann constant, $T_{i}$ and $T_{e}$ are, respectively, the average temperatures of ions and electrons, $m_{i}$ is the mass of an ion,  $n_{d0}$ is the dust number density, $Z_{d}$ is the number of negative unit charges  residing on dust grain surface, and $e$ is the charge of an electron.

The above equations are supplemented by nonthermally distributed electrons as prescribed by Cairns \textit{et al} \cite{Cairns95} for the electron species. Nonthermal distribution of any lighter species of particles (as prescribed by Cairns \textit{et al} \cite{Cairns95} for the electron species) can be regarded as a modified Boltzmannian distribution, which has the property that the number of particles in phase space in the neighborhood of the point $v=0$ is much smaller than the number of particles in phase space in the neighborhood of the point $v=0$ for the case of Boltzmann distribution, where $v$ is the velocity of the particle in phase space. Under the above mentioned normalization of the dependent and independent variables, the normalized number density of nonthermal electrons can be written as
\begin{eqnarray}\label{dist func electron non}
n_{e}=(1-\beta_{1} \phi + \beta_{1} \phi^{2})e^{\phi},
\end{eqnarray}
where
\begin{eqnarray}\label{beta1}
\beta_{1} = \frac{4 \alpha_{1}}{1+3 \alpha_{1}},
\end{eqnarray}
with $\alpha_{1} \geq 0$. Here $\beta_{1} (\alpha_{1})$ is the parameter associated with nonthermal distribution of electrons and this parameter determines the proportion of fast energetic electrons. From (\ref{beta1}) and the inequality $\alpha_{1} \geq 0$, it can be easily checked that the nonthermal parameter $\beta_{1}$ is restricted by the following inequality: $0 \leq \beta_{1} < 4/3$. However we cannot take the whole region of $\beta_{1}$ ($0\leq \beta_{1}< 4/3$). Plotting the nonthermal velocity distribution of electrons against its velocity ($v$) in phase space, it can be easily shown that the number of electrons in phase space in the neighborhood of the point $v=0$ decreases with increasing $\beta_{1}$ and the number of electrons in phase space in the neighborhood of the point $v=0$ is almost zero when $\beta_{1}\rightarrow 4/3$. Therefore, for increasing $\beta_{1}$ distribution function develops wings, which become stronger as $\beta_{1}$ increases, and at the same time the center density in phase space drops, the latter as a result of the normalization of the area under the integral. Consequently, we should not take values of $\beta_{1} > 4/7$ since that stage might stretch the credibility of the Cairns model too far \cite{Verheest08}. So, here we consider the effective range of $\beta_{1}$ as follows: $0\leq \beta_{1} \leq \beta_{1T}$, where $\beta_{1T} = 4/7\approx 0.571429$.

The charge neutrality condition,
\begin{equation}\label{charge neutrality without mu}
Z_{d} n_{d0}+n_{e0}=n_{i0},
\end{equation}
can be written as
\begin{equation}\label{charge neutrality with mu}
\frac{Z_{d} n_{d0}}{n_{i0}} = 1 -\mu,
\end{equation}
where $\mu = n_{e0}/n_{i0}$ and consequently, the Poisson equation (\ref{Poisson eq without mu}) assumes the following form:
\begin{eqnarray}\label{Poisson eq}
\phi_{xx} = \mu n_{e}- n_{i} + (1-\mu).
\end{eqnarray}
We note from (\ref{charge neutrality with mu}) that $1-\mu$ must be greater than zero, i.e., $0 < \mu < 1$. When $\mu \rightarrow 1$, the effect of negatively charged dust grains on DIA wave is negligible and so, we restrict $\mu$ by the inequality $0 < \mu \leq \mu_{T}$, where $\mu_{T}$ is strictly less than 1.
\section{\label{sec:Energy int}Energy integral and the Sagdeev potential}
To study the arbitrary amplitude time independent DIASWs and DIADLs we make all the dependent variables depend only on a single variable $\xi=x-Mt$, where the Mach number $M$ is normalized by $c_{i}$. Thus in the steady state, (\ref{continuity of ions}) - (\ref{pressure eq}) and (\ref{Poisson eq}) can be written as
\begin{eqnarray}\label{continuity of ions steady state}
-M n_{i,~\xi} + (n_{i}u_{i})_{\xi} = 0,
\end{eqnarray}
\begin{eqnarray}\label{motion of ions steady state}
-M u_{i,~\xi} + u_{i} u_{i,~\xi} = -\phi_{\xi}-\alpha n_{i}^{-1}
p_{i,~\xi},
\end{eqnarray}
\begin{eqnarray}\label{pressure eq steady state}
-M p_{i,~\xi} + u_{i} p_{i,~\xi} + 3 p_{i} u_{i,~\xi} = 0,
\end{eqnarray}
\begin{eqnarray}\label{Poisson eq steady state}
\phi_{\xi \xi} = \mu n_{e}- n_{i} + (1-\mu).
\end{eqnarray}
Using the boundary conditions,
\begin{eqnarray}\label{boundary cond}
&& n_{i}\rightarrow 1, p_{i}\rightarrow 1, u_{i}\rightarrow 0,\nonumber \\ 
&& \phi \rightarrow 0, \phi_{\xi}\rightarrow 0 ~\mbox{as}~ |\xi| \rightarrow
\infty,
\end{eqnarray}
and solving (\ref{continuity of ions steady state}) - (\ref{pressure eq steady state}), we get a quadratic equation for $n_{i}^{2}$ and following the same argument as given in Das \textit{et al.} \cite{ADas} to find the expression of dust density with exact bounds, we get the following expression of $n_{i}$.
\begin{eqnarray}\label{n_i with Phi_M Psi_M 2}
    n_{i} = \frac{\sqrt{2}M}{\sqrt{\Psi_{M}-\phi}+\sqrt{\Phi_{M}-\phi}}.
\end{eqnarray}
where
\begin{eqnarray}\label{Psi_M Phi_M}
\Psi_{M}=\frac{(M-\sqrt{3 \alpha})^{2}}{2},
\Phi_{M}=\frac{(M+\sqrt{3 \alpha})^{2}}{2}.
\end{eqnarray}
From (\ref{n_i with Phi_M Psi_M 2}), we see that this equation gives both theoretically and  numerically correct expression of $n_{i}$ even when $\alpha = 0$ if $\phi\leq \Psi_{M}$.

Now integrating (\ref{Poisson eq steady state}) with respect to $\phi$ and using the boundary conditions (\ref{boundary cond}), we get the following energy integral with $V(\phi)$ as Sagdeev potential or pseudo-potential.
\begin{eqnarray}\label{energy int}
\frac{1}{2} \bigg(\frac{d\phi}{d\xi}\bigg)^{2}+V(\phi) = 0,
\end{eqnarray}
where
\begin{eqnarray}\label{v(phi)}
    V(\phi)\equiv V(M,\phi) =  V_{i}-\mu V_{e}  - (1-\mu) \phi,
\end{eqnarray}
\begin{eqnarray}\label{V_e non}
    V_{e} = (1+3\beta_{1}-3 \beta_{1}\phi+\beta_{1}\phi^{2})e^{\phi}-(1+3\beta_{1}),
\end{eqnarray}
\begin{eqnarray}\label{V_i}
    V_{i} = M^{2} + \alpha -n_{i} (M^{2}+3\alpha-2 \phi-2
    \alpha n_{i}^{2}).
\end{eqnarray}
Here, $V(M,\phi)$ is same as $V(\phi)$, i.e., the Mach number $M$ is omitted from the notation $V(M,\phi)$ when no particular emphasis is put upon it.
\section{\label{sec:physics}Physical interpretation of the energy integral}
The energy integral (\ref{energy int}) can be regarded as the one-dimensional motion of a particle of unit mass whose position is $\phi$ at time $\xi$ with velocity $d\phi/d\xi$ in a potential well $V(\phi)$. The first term in the energy integral (\ref{energy int}) can be regarded as the kinetic energy of a particle of unit mass at position $\phi$ and time $\xi$ whereas $V(\phi)$ is the potential energy at that instant. Since kinetic energy is always a non-negative quantity, $V(\phi) \leq 0$ for the entire motion, i.e., zero is the maximum value for $V(\phi)$. Again from (\ref{energy int}), we find $d^{2} \phi/d \xi^{2} + V'(\phi) = 0$, i.e. the force acting on the particle of unit mass at the position $\phi$ is $-V'(\phi)$, where ``$~'~$'' indicates a derivative with respect to $\phi$. Now, it can be easily checked that $V(0) = V'(0) = 0$, and consequently, the particle is in equilibrium at $\phi=0$ because the velocity as well as the force acting on the particle at $\phi=0$ are simultaneously zero. Now if $\phi = 0$ can be made an unstable position of equilibrium, the energy integral can be interpreted as the motion of an oscillatory particle if $V(\phi_{m}) = 0$ for some $\phi_{m} \neq 0$, i.e., if the particle is slightly displaced from its unstable position of equilibrium then it moves away from its unstable position of equilibrium and it continues its motion until its velocity is equal to zero, i.e., until $\phi$ takes the value $\phi_{m}$. Now the force acting on the particle of unit mass at position $\phi = \phi_{m}$ is $-V'(\phi_{m})$. For $\phi_{m}<0$, the force acting on the particle at the point $\phi=\phi_{m}$ is directed towards the point $\phi=0$ if $-V'(\phi_{m})>0$, i.e., if $V'(\phi_{m})<0$. On the other hand, for $\phi_{m}>0$, the force acting on the particle at the point $\phi=\phi_{m}$ is directed towards the point $\phi=0$ if $-V'(\phi_{m})<0$, i.e., if $V'(\phi_{m})>0$. Therefore, if $V'(\phi_{m})>0$ (for the positive potential side ) or if $V'(\phi_{m})<0$ (for the negative potential side ) then the particle reflects back again to $\phi = 0$. Again, if $V(\phi_{m})=V'(\phi_{m})=0$ then the velocity $d \phi/d\xi$ as well as the force $d^{2} \phi/d \xi^{2}$ both are equal to zero at $\phi = \phi_{m}$. Consequently, if the particle is slightly displaced from its unstable position of equilibrium ($\phi = 0$) it moves away from $\phi = 0$ and it continues its motion until the velocity is equal to zero, i.e., until $\phi$ takes the value $\phi = \phi_{m}$. However it cannot be reflected back again at $\phi = 0$ as the velocity and the force acting on the particle at $\phi = \phi_{m}$ vanish simultaneously. Actually, if $V'(\phi_{m})>0$ (for $\phi_{m}>0$) or if $V'(\phi_{m})<0$ (for $\phi_{m}<0$) the particle takes an infinite long time to move away from the unstable position of equilibrium. After that it continues its motion until $\phi$ takes the value $\phi_{m}$ and again it takes an infinite long time to come back its unstable position of equilibrium. Therefore, for the existence of a positive (negative) potential solitary wave solution of the energy integral (\ref{energy int}), we must have the following: (a) $\phi=0$ is the position of unstable equilibrium of the particle, (b) $V(\phi_{m}) = 0$, $V'(\phi_{m}) > 0$ $(V'(\phi_{m}) < 0)$ for some $\phi_{m} > 0$ $(\phi_{m} < 0)$, which is nothing but the condition for oscillation of the particle within the interval $\min\{0,\phi_{m}\}<\phi<\max\{0,\phi_{m}\}$  and (c) $V(\phi) < 0$ for all $0 <\phi < \phi_{m}$ $(\phi_{m} < \phi < 0$), which is the condition to define the energy integral (\ref{energy int}) within the interval $\min\{0,\phi_{m}\}<\phi<\max\{0,\phi_{m}\}$. For the existence of a positive (negative) potential DL solution of the energy integral (\ref{energy int}), the conditions (a) and (c) remain unchanged but here (b) has been modified in such a way that the particle cannot be reflected again at $\phi = 0$, i.e., the condition (b) assumes the following form: $V(\phi_{m}) = V'(\phi_{m}) = 0$, $V''(\phi_{m}) < 0$ for some $\phi_{m} > 0$ $(\phi_{m} < 0$).
\section{\label{sec:mach number}Lower and Upper bounds of Mach number for the existence of solitary waves and double layers}
For the existence of solitary structures, we must have $V(0)=V'(0)=0$ and $V''(0) < 0$. Now it can be easily verified that the first two conditions, i.e., $V(0)=0$ and $V'(0)=0$ are trivially satisfied whereas, the condition $V''(0) < 0$ gives $M>M_{c}$, where $M_{c}$ is given by the following equation.
\begin{eqnarray}\label{Mc nonthermal}
    M_{c}^{2} = 3 \alpha+\frac{1}{\mu (1-\beta_{1})}.
\end{eqnarray}
Now, for $M_{c}$ to be real and positive, we must have $\mu>0$ and $0\leq \beta_{1}<1$. As the effective range of $\beta_{1}$ is $0\leq \beta_{1} \leq \beta_{1T}$, where $\beta_{1T} = 4/7\approx 0.571429$, $M_{c}$ is well-defined as a real positive quantity for all $0<\mu\leq \mu_{T}$ and $0\leq \beta_{1} \leq \beta_{1T}$.
\subsection{\label{subsec:positive}Upper bounds of the Mach number for the existence of PPSWs}
Consider the existence of a PPSW for some value of $M > M_{c}$. Therefore, there exists a $\phi_{m}>0$ such that
\begin{equation}\label{ppsw}
   \left .\begin{array}{ccc}
     V(\phi) < 0 & ~\mbox{for all}~& 0 < \phi < \phi_{m}, \\
     V(\phi_{m}) = 0 & ,&  V'(\phi_{m}) > 0.
   \end{array}\right \}
\end{equation}
Now as $V(\phi)$ is real for $\phi \leq \Psi_{M}$ we must have $\phi_{m} \leq \Psi_{M}$, otherwise $V(\phi_{m})$ is not a real quantity. Therefore,
\begin{equation}
    V(\phi) < 0 ~~\mbox{for all}~~0 < \phi < ( \phi_{m}\leq)\Psi_{M},
\end{equation}
defines a large amplitude PPSW, which is in conformity with (\ref{ppsw}), if $V(\Psi_{M}) = 0$ and $V'(\Psi_{M}) > 0$.

Again let $M_{max}$ be the maximum value of $M$ up to which solitary wave solution can exist. As $\Psi_{M}$ increases with $M$ then $\Psi_{M} \leq \Psi_{M_{max}}$. Therefore,
\begin{equation*}
    V(\phi) < 0 ~~\mbox{for all}~~0 < \phi <
    (\phi_{m} \leq \Psi_{M} \leq )\Psi_{M_{max}} ,
\end{equation*}
defines the largest amplitude PPSW if $V(\Psi_{M_{max}}) = 0$ and $V'(\Psi_{M_{max}}) > 0$. Therefore, for the existence of the PPSWs, the Mach number $M$ is restricted by the following inequality: $M_{c}<M\leq M_{max}$, where $M_{max}$ is the largest positive root of the equation $V(\Psi_{M})= 0$ subject to the condition $V(\Psi_{M})\geq 0$ $\mbox{for all}$ $M \leq M_{max}$. Now if at $M=M_{d}$ ($M_{c}<M_{d}< M_{max}$), one can get a PPDL solution of the energy integral (\ref{energy int}) then for the existence of the PPSWs, the Mach number $M$ is restricted by the inequality: $M_{c}<M< M_{d}$ and also $M_{d}<M\leq M_{max}$. On the other hand if $M_{d}= M_{max}$, then for the existence of PPSWs, the Mach number $M$ is restricted by the inequality: $M_{c}<M< M_{d}$.
\subsection{\label{subsec:negative}Upper bounds of the Mach number for the existence of NPSWs}
We have seen earlier that $V(\phi)$ is real if $\phi \leq \Psi_{M} $, where $\Psi_{M}$ is strictly positive. For NPSWs or NPDLs, we have
\begin{eqnarray}
    V(\phi)<0 ~~\mbox{for all }~~\phi_{m}<\phi<0,
\end{eqnarray}
along with the other conditions stated in section \ref{sec:physics} for the existence of NPSWs or NPDLs. As $\Psi_{M}$ is strictly positive and for NPSWs or NPDLs, $\phi <0$, the condition $\phi < \Psi_{M}$ is automatically satisfied and consequently, for these two cases $V(\phi)$ is well defined for all $\phi <0$ without imposing extra condition. Since there is no such restriction on $\phi$, we cannot use the same definition as in the case of PPSWs to find the upper bound of Mach numbers for the existence of NPSWs. For the case of NPSWs, to find an upper limit or upper bound of $M$, up to which NPSW can exist, we shall first of all find a value $M_{D}$ of $M$ for which energy integral (\ref{energy int}) gives a NPDL solution at $M=M_{D}$ with amplitude $\phi = \phi_{D}$. Now if at $M=M_{D}$, $\phi=\phi_{D}$ is the only root (double root) of the equation $V(\phi)\equiv V(M, \phi) = 0$, i.e., $V(M_{D}, \phi_{D})=0$ and $V'(M_{D}, \phi_{D})=0$, then the NPDL solution is the ultimate solution of the
energy integral (\ref{energy int}) and in this case, no NPSW solution can be obtained if $M>M_{D}$, i.e., for the occurrence of NPSWs, the Mach number $M$ is restricted by the inequality $M_{c}<M<M_{D}$. 

On the other hand if there exists an inaccessible simple root $\phi_{D1}$ of $\phi$ such that $\phi_{D1}<\phi_{D}$, i.e., $V(M_{D}, \phi_{D_{1}})=0$ along with $V(M_{D}, \phi_{D})=0$, $V'(M_{D}, \phi_{D})=0$ and $\phi_{D1}<\phi_{D}$, then there exists a NPSW solution of the energy integral (\ref{energy int}) for at least one value of $M>M_{D}$. Hence in the later case double layer solution is unable to restrict the occurrence of NPSWs for $M>M_{D}$. From this consideration, it is also clear that the occurrence of NPSWs is restricted by $M_{c}<M<M_{D}$, provided that there exists one and only one double root of the equation $V(M_{D},\phi)=0$ for the unknown $\phi \neq 0$, otherwise NPSWs can exist for all $M > M_{D}$.

Therefore, the double layer solution of the energy integral (\ref{energy int}) plays an important role to determine the upper bound of the Mach number for existence of either PPSWs or NPSWs. For the present problem, PPSW exists if $M_{c}<M\leq M_{max}$ but from the discussion of \sref{subsec:positive} it is not clear whether the energy integral (\ref{energy int}) provides a PPDL solution for some $M=M_{d}$ such that $M_{c}<M_{d}\leq M_{max}$. Similarly, from the discussion of \ref{subsec:negative} it is not clear whether the energy integral (\ref{energy int}) provides a NPDL solution for some $M=M_{D}$. So, in the next subsection, we shall analytically investigate the existence of double layer solution of the energy integral (\ref{energy int}).
\subsection{\label{subsec:dl}Analytical study of the Double layer solution of the Energy Integral}
From the condition (b) as given in \sref{sec:physics} for the existence of double layer solution of the energy integral (\ref{energy int}), we must have a non-zero $\phi$ ($\phi\neq 0$) such that following conditions are simultaneously satisfied:
\begin{eqnarray}\label{dlb1}
    V(\phi)=0,
    V'(\phi)=0,
    V''(\phi)<0.
\end{eqnarray}
Using (\ref{v(phi)}) - (\ref{V_i}), the first equation, the second equation and the third inequality of (\ref{dlb1}) can be written, respectively, as
\begin{eqnarray}\label{redlb1}
    V(\phi)\equiv (1-n_{i})M^{2} + \alpha -n_{i} (3\alpha-2 \phi-2
    \alpha n_{i}^{2})\nonumber \\-S=0,
\end{eqnarray}
\begin{eqnarray}\label{redlb2}
    V'(\phi)\equiv n_{i}-\frac{dS}{d \phi}=0,
\end{eqnarray}
\begin{eqnarray}\label{redlb3}
    V''(\phi)\equiv \frac{d}{d \phi}\bigg(n_{i}-\frac{dS}{d \phi}\bigg)<0,
\end{eqnarray}
where
\begin{eqnarray}\label{S}
    S=\mu V_{e}+(1-\mu)\phi.
\end{eqnarray}
Eliminating $n_{i}$ from (\ref{redlb1}) and (\ref{redlb2}), we get
\begin{eqnarray}\label{Mdl}
    \mu(1-n_{e})M^{2} = S + \frac{dS}{d \phi} \bigg[3\alpha-2 \phi-2
    \alpha \bigg(\frac{dS}{d \phi}\bigg)^{2}\bigg]\nonumber \\- \alpha.
\end{eqnarray}
It can be easily checked that $\phi=0$ if and only if $n_{e}=1$, and consequently, for non-zero $\phi$, (\ref{Mdl}) can be written as
\begin{eqnarray}\label{reMdl}
    M^{2} = h(\phi),
\end{eqnarray}
where
\begin{eqnarray}\label{hphi}
    h(\phi) = \frac{S + \frac{dS}{d \phi} \bigg[3\alpha-2 \phi-2
    \alpha \bigg(\frac{dS}{d \phi}\bigg)^{2}\bigg]- \alpha}{\mu(1-n_{e})}.
\end{eqnarray}
Using (\ref{reMdl}), from (\ref{redlb2}) and (\ref{redlb3}), we, respectively, get
\begin{eqnarray}\label{dlamp}
\eta(\phi)\equiv\frac{\sqrt{2h(\phi)}}{\sqrt{g_{+}(\phi)-\phi}+\sqrt{g_{-}(\phi)-\phi}}-\frac{dS}{d \phi}=0,
\end{eqnarray}
\begin{eqnarray}\label{dlcon}
    \frac{d\eta}{d \phi}<0,
\end{eqnarray}
where
\begin{eqnarray}\label{gpm}
    g_{\pm}(\phi)=\frac{1}{2}(\sqrt{h(\phi)}\pm \sqrt{3\alpha})^{2}.
\end{eqnarray}
Now suppose $\phi=\phi_{dl}\neq 0$ solves (\ref{dlamp}), then the double layer solution of the energy integral (\ref{energy int}) exists at $M=M_{dl}=\sqrt{h(\phi_{dl})}$ having amplitude $|\phi_{dl}|$ if the following conditions are satisfied:
\begin{eqnarray}\label{con1}
    h(\phi_{dl})-M_{c}^{2}>0,
\end{eqnarray}
\begin{eqnarray}\label{con2}
    g_{-}(\phi_{dl})-\phi_{dl}\geq 0 ,
\end{eqnarray}
\begin{eqnarray}\label{con3}
    \frac{d \eta}{d \phi}\bigg|_{\phi=\phi_{dl}}<0.
\end{eqnarray}
To derive the condition (\ref{con2}), we have used the following restriction on $\phi$ : $\phi \leq \Psi_{M}=\frac{1}{2}(M-\sqrt{3\alpha})^{2}$. Again the condition (\ref{con3}) states that the function $\eta(\phi)$ decreases in a neighbourhood of $\phi_{dl}$ and $\eta(\phi_{dl})=0$. Therefore if $\phi_{dl}>0$, $\eta(\phi)$ changes sign from positive to negative when it crosses the point $\phi_{dl}$ from left to right. However, for $\phi_{dl}<0$, $\eta(\phi)$ changes sign from positive to negative when it crosses the point $\phi_{dl}$ from right to left. In any case, $\eta(\phi)$ vanishes at $\phi = \phi_{dl}$.
\subsection{\label{algo}Algorithm to find negative (positive) potential double layer solution}
From this theoretical discussion, it is simple to make a numerical scheme to test whether the energy integral provides a double layer solution. The simple algorithm for the existence of double layer solution can be written as follows.\\
\textbf{Step - 1:}	Find $h(\phi)$ defined by (\ref{hphi}) for $\phi < 0$ ($\phi > 0$). If $h(\phi)$ is negative for all $\phi < 0$ ($\phi > 0$) then the system does not support negative (positive) potential double layer. Else follow the next step.\\
\textbf{Step - 2:}	If $\sqrt{g_{+}(\phi)-\phi}+\sqrt{g_{-}(\phi)-\phi}$ be real for $\phi < 0$ ($\phi > 0$) then go to next step. Otherwise, the system does not support negative (positive) potential double layer.\\
\textbf{Step - 3:} Set up a numerical scheme to find all possible real negative (positive) roots of $\phi$ of (\ref{dlamp}) for all admissible values of the parameters. To find these roots of $\phi$ at some fixed values of the parameters, let $\phi$ free and take all those $\phi$'s where the function $\eta(\phi)$ changes its sign from positive to negative when it crosses the point $\phi$ from right to left (left to right) and denote the set of all these $\phi$'s as $\mathcal{D_{-}}$ ($\mathcal{D_{+}}$). Obviously, there may exist other roots of $\phi$ of (\ref{dlamp}) for unknown $\phi$ such that $\eta(\phi)$ changes its sign from negative to positive, however, the theoretical discussions suggest that these roots are beyond the scope of the present numerical scheme.\\
\textbf{Step - 4:} Using (\ref{reMdl}) find the Mach numbers $M$ at each $\phi$ of $\mathcal{D_{-}}$ ($\mathcal{D_{+}}$).\\
\textbf{Step - 5:} Find the largest (smallest) value $\phi_{D}$ ($\phi_{d}$) of $\phi$ from $\mathcal{D_{-}}$ ($\mathcal{D_{+}}$) such that conditions (\ref{con1}) - (\ref{con2}) hold good. Then, obviously $\phi_{D}$ ($\phi_{d}$) is the amplitude of NPDL (PPDL) solution. Next use (\ref{reMdl}) to obtain the Mach number $M_{D}$ ($M_{d}$) corresponding to NPDL (PPDL) solution having amplitude $|\phi_{D}|$ ($\phi_{d}$).

With the help of this algorithm, we want to investigate numerically the following facts for the present system:\\
(i) Whether the present system supports PPDL and/or NPDL solutions. (ii) Whether the double layer solution, if exists, can restrict the occurrence of all solitons of same polarity, i.e., whether the double layer solution is the ultimate solution of all solitons of same polarity of the present system. In this connection, it is important to remember that for any double layer solution, there must exists at least one sequence of solitons of same polarity converging to the double layer solution, i.e., the amplitude of any double layer solution acts as an exact upper bound of the amplitudes of at least one sequence of solitary waves of same polarity.

\begin{figure}
\begin{center}
  \includegraphics{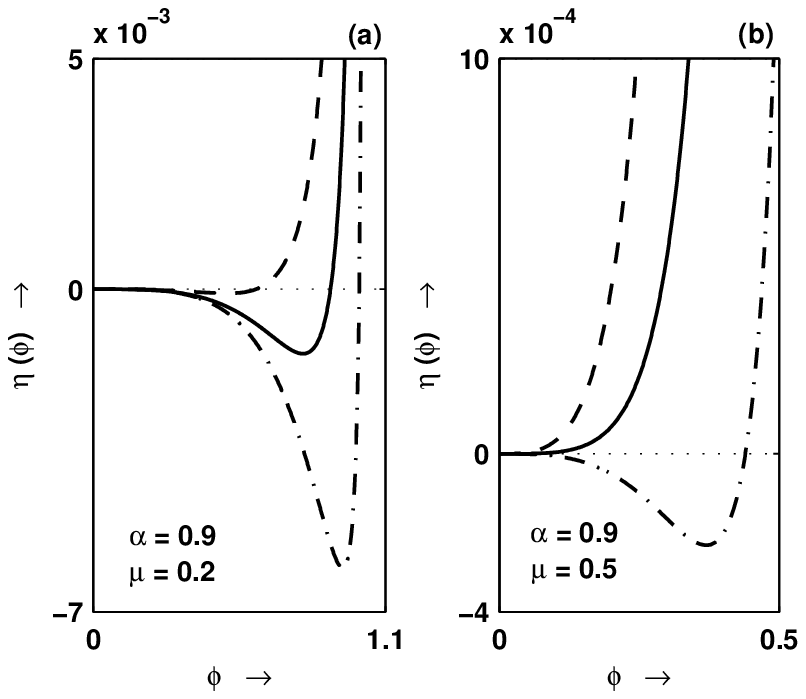}
  \caption{\label{fig:dl pos non} $\eta(\phi)$ is plotted against $\phi$ for $\alpha = 0.9$ with $\mu = 0.2$ in (a) and $\mu = 0.5$ in (b). In each of (a) and (b), three curves have been drawn corresponding to three different values of $\beta_{1}$, viz., $\beta_{1}$ are 0.2[- - -], 0.4[---] and 0.6[$- \cdot -$].}
\end{center}
\end{figure}

To investigate the existence of PPDL, $\eta(\phi)$ is plotted against $\phi$ for $\phi>0$ in \fref{fig:dl pos non} for different values of $\mu$ and $\beta_{1}$. \Fref{fig:dl pos non}(a) and \fref{fig:dl pos non}(b) show that $\eta(\phi)$ is either remain positive throughout $\phi > 0$ or changes sign from negative to positive when it crosses positive $\phi$ axis from left to right. In \fref{fig:dl pos non}(a), $\eta(\phi)$ corresponding to $\beta_{1} = 0.4$ changes sign from negative to positive when it crosses positive $\phi$ axis from left to right. Similar facts have been observed in \fref{fig:dl pos non}(a) for $\eta(\phi)$ corresponding to $\beta_{1} = 0.6$ and in \fref{fig:dl pos non}(b) for $\eta(\phi)$ corresponding to $\beta_{1} = 0.6$. Consequently, there does not exist any PPDL solution. For any admissible values of the parameters involved in the system, it can be easily verified that the system does not support any PPDL solution with the help of plotting $\eta(\phi)$ against $\phi$ for $\phi>0$. Since the system does not support any PPDL solution, we can conclude that $M_{max}$ is the only upper bound of $M$ for the occurrence of PPSWs. Thus PPSWs exist whenever $M_{c}<M \leq M_{max}$.

\begin{figure}
\begin{center}
  \includegraphics{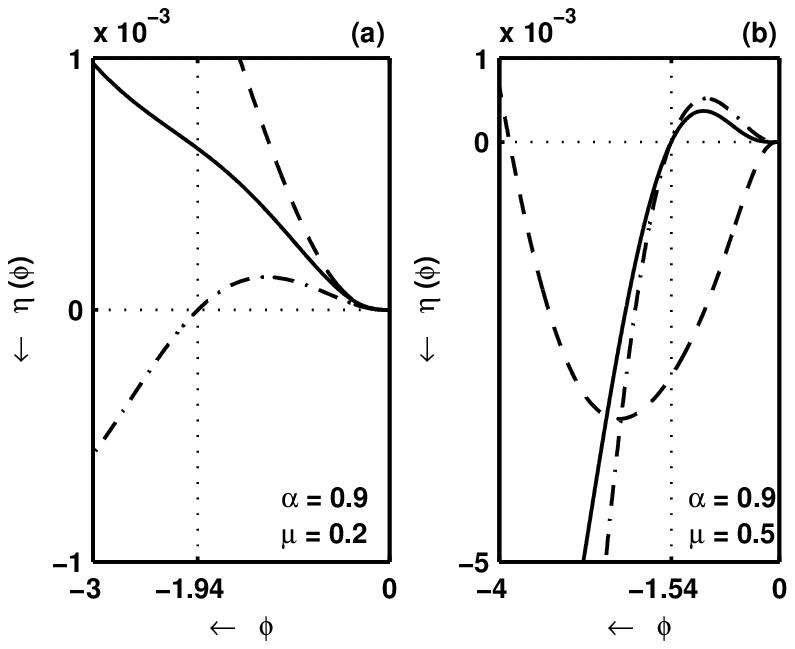}
  \caption{\label{fig:dl neg non} $\eta(\phi)$ is plotted against $\phi$ for $\alpha = 0.9$ with $\mu = 0.2$ in (a) and $\mu = 0.5$ in (b). In each of (a) and (b), three curves have been drawn corresponding to three different values of $\beta_{1}$, viz., $\beta_{1}$ are 0.2[- - -], 0.4[---] and 0.6[$- \cdot -$].}
\end{center}
\end{figure}

To investigate the existence of NPDL, $\eta(\phi)$ is plotted against $\phi$ for $\phi<0$ in \fref{fig:dl neg non}. In \fref{fig:dl neg non}(a), $\eta(\phi)$ corresponding to $\beta_{1} = 0.2$ and $\beta_{1} = 0.4$ remain positive throughout $\phi < 0$ and thus there does not exist NPDL solution for $\beta_{1} = 0.2$ and also for $\beta_{1} = 0.4$ with $\alpha=0.9$, $\mu = 0.2$. It can be easily checked that there does not exist any NPDL solution for $0 \leq \beta_{1} \leq 0.4$ with $\alpha=0.9$, $\mu = 0.2$ by simply drawing $\eta(\phi)$ against $\phi$. However $\eta(\phi)$ corresponding to $\beta_{1} = 0.6$ changes sign from positive to negative when it crosses negative $\phi$ axis from right to left and fulfill all the conditions of our algorithm for the existence of NPDL solution. A similar interpretation of \fref{fig:dl neg non}(b) shows that NPDLs exist for $\beta_{1} \geq 0.4$ with $\alpha=0.9$, $\mu = 0.5$. Thus for larger values of $\mu$, NPDL solution is expected at lower values of $\beta_{1}$. Therefore, for denser electrons in the background plasma, we require less amount of fast energetic electrons to get NPDL. In other words, electron density depletion restrict the occurrence of NPDL. 

Still it is not clear whether the existence of NPDL can restrict all NPSWs of the present system, or in other words, whether $M_{D}$ is the upper limit of $M$ for the existence of all NPSWs of the present system. From \fref{fig:dl neg non}, one can find three set of values of the parameters $\alpha$, $\mu$ and $\beta_{1}$ such that NPDL exist. It is easy to find $\phi_{D}$, at which $\eta(\phi)$ changes sign from positive to negative when it crosses negative $\phi$ axis from right to left and using (\ref{reMdl}) we can find three values of $M_{D}$ corresponding to three different values of $\phi_{D}$ for three set of values of the parameters $\alpha$, $\mu$ and $\beta_{1}$ as shown in \fref{fig:dl neg non}. For clarity we have tabulate these values in \tref{tab:table1}. Our algorithm will be verified if we can confirm the occurrence of NPDL solutions at those values of parameters.
\begin{table}
\begin{center}
\caption{\label{tab:table1}The values of $\phi_{D}$ and the corresponding values $M_{D}$ calculated from (\ref{reMdl}) have been tabulated for different values of $\alpha$, $\mu$ and $\beta_{1}$, which have been used in \fref{fig:dl neg non}.}
\begin{indented}
\item[]\begin{tabular}{@{}cccccc}
\br
 &$\alpha$&$\mu$&$\beta_{1}$&$M_{D}$&$\phi_{D}$\\
\mr
$V_{1}$:&0.9&0.2&0.6&5.10476181&-1.9397924\\
$V_{2}$:&0.9&0.5&0.4&2.49995351&-1.5418283\\
$V_{3}$:&0.9&0.5&0.6&3.25924103&-1.5492367\\
\br
\end{tabular}
\end{indented}
\end{center}
\end{table}

\begin{figure}
\begin{center}
  \includegraphics{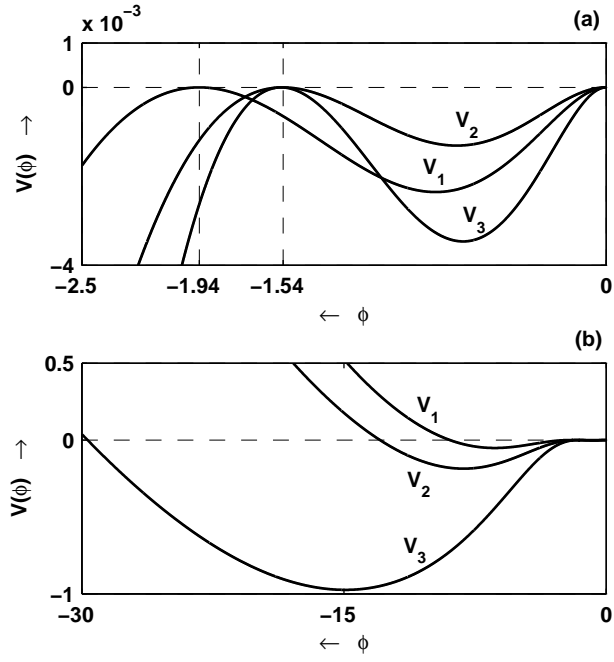}
  \caption{\label{fig:verify dl} $V(\phi)$ is plotted against $\phi$ for three set of values $V_{1}$, $V_{2}$ and $V_{3}$ presented in \tref{tab:table1}. In figure (a) all the curves ($V_{1}$, $V_{2}$ and $V_{3}$) show the existence of NPSW solutions. In figure (b) the same curves have been plotted in an interval of $\phi$ close to $\phi = 0$.}
\end{center}
\end{figure}

In \fref{fig:verify dl}(a), $V(\phi)$ is plotted against $\phi$ for three set of values (denoted as $V_{1}$, $V_{2}$ and $V_{3}$ in \tref{tab:table1}) of $\alpha$, $\mu$, $\beta_{1}$ and $M_{D}$. Our aim is to show the amplitudes obtained from \fref{fig:verify dl}(a) are exactly the same as obtained from \tref{tab:table1}. Each of the curves in \fref{fig:verify dl}(a) shows the existence of a NPDL solution. Moreover, the amplitudes of these double layers are exactly the same as obtained in \tref{tab:table1}. Hence our algorithm regarding the double layer solution is correct. From \tref{tab:table1}, we have some typical observations regarding the amplitude of NPDLs. For fixed $\alpha$ and $\mu$ the amplitude of double layer increases with $\beta_{1}$. Again for fixed values of $\alpha$ and $\beta_{1}$, the amplitude of double layer decreases with increasing $\mu$. In other words, NPDL gets stronger with fast energetic electrons. However, for any fixed non-zero value of $\beta_{1}$ and for any value of $\alpha$, one can get stronger NPDL by adding more electrons on the dust grain surface. 

\begin{figure}
\begin{center}
  \includegraphics{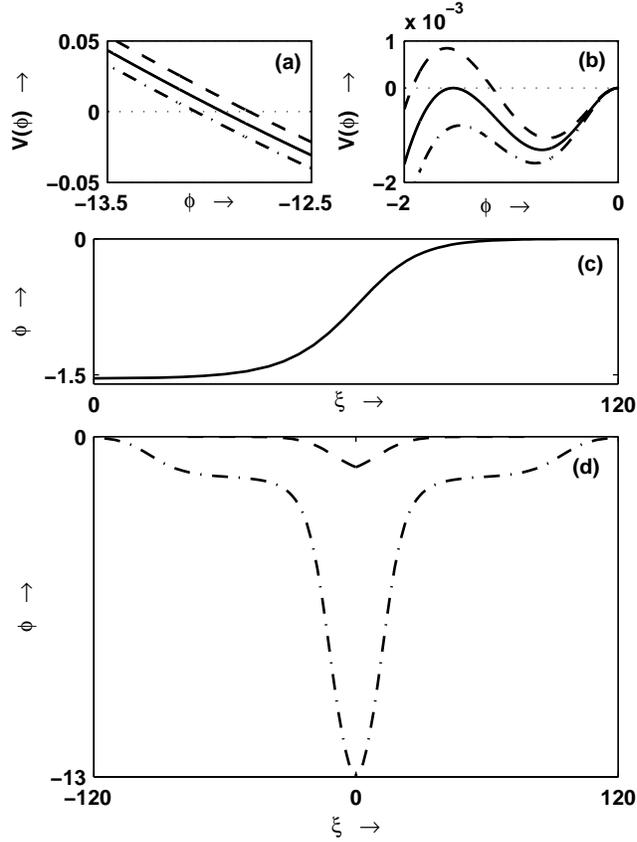}
  \caption{\label{fig:solitary and dl}$V(\phi)$ is plotted against $\phi$ for three different values of $M$, viz., $M_{D}$[---], $M_{D}-0.005$[- - -] and $M_{D}+0.005$[$- \cdot -$] in (a) and (b). In (b) we have shown the region of $\phi$ from $\phi = -2$ to $\phi = 0$, whereas in (a) a region of $\phi$ from $\phi = -13.5$ to $\phi = -12.5$ has been shown. The NPDL profile corresponding to $M=M_{D}$ has been shown in (c), whereas the profiles of NPSWs corresponding to $M=M_{D}-0.005$ and $M=M_{D}+0.005$ have been drawn in (d). Profiles in (d) show a finite jump between the amplitudes of solitary wave by going from $M=M_{D}-0.005$ to $M=M_{D}+0.005$. In all these four figures we have used $\alpha = 0.9$, $\mu = 0.5$ and $\beta_{1} = 0.4$.}
\end{center}
\end{figure}

Now we are in a position to investigate whether the NPDL solution can restrict the occurrence of all NPSWs of the present system. For this purpose, we explore \fref{fig:verify dl}(a) beyond $\phi_{D}$ and obtain \fref{fig:verify dl}(b). In this figure, we have drawn the same set of curves as in \fref{fig:verify dl}(a) with an exception that we have extended the range of $\phi$ axis far away from $\phi = 0$. We see that after making a double root at $\phi = \phi_{D}$, there exists an $\phi_{D1}<\phi_{D}$ such that $V(M_{D},\phi_{D1}) = 0$. Thus according to our theoretical discussions in \sref{subsec:negative}, there exists a $M>M_{D}$ such that NPSW exists. Therefore, the NPDL solution cannot restrict the occurrence all NPSWs of the present system and consequently, $M_{D}$ cannot act as an upper bound of $M$ for the existence of all NPSWs of the present system. Actually, the present system supports very large amplitude NPSW for all $M>M_{D}$. To justify this fact, in \fref{fig:solitary and dl}(a) and \fref{fig:solitary and dl}(b), $V(\phi)$ is plotted against $\phi$ for three different values of $M$, viz., $M_{D}$, $M_{D}-0.005$ and $M_{D}+0.005$. \Fref{fig:solitary and dl}(b) shows that at $M=M_{D}$, there exists a NPDL of amplitude $|\phi_{D}|$ whereas, for $M=M_{D}-0.005$, there exists a NPSW of amplitude less than $|\phi_{D}|$. However, \fref{fig:solitary and dl}(b) shows that the equation $V(M_{D}+0.005,\phi) = 0$ has no real root of $\phi$ in the neighborhood of $\phi=\phi_{D}$. From \fref{fig:solitary and dl}(a), we see that $V(M,\phi)$ again vanishes at $\phi = \phi_{1}$, $\phi = \phi_{2}$ and $\phi=\phi_{3}$, respectively, for $M=M_{D}-0.005$,  $M=M_{D}$, and $M=M_{D}+0.005$. However, the roots $\phi = \phi_{1}$ and $\phi = \phi_{2}$ of $V(M,\phi)=0$ corresponding to $M=M_{D}-0.005$ and $M=M_{D}$ are unable to give any solitary wave solution, whereas the root $\phi=\phi_{3}$ of $V(M_{D}+0.005,\phi)=0$ gives a NPSW of amplitude much greater than that of NPDL at $M=M_{D}$ as well as NPSW at $M=M_{D}-0.005$. Thus there is a finite jump in amplitudes between two NPSWs at $M=M_{D}-0.005$ and at $M=M_{D}+0.005$ separated by the NPDL at $M=M_{D}$. This is not a new result, the same result has also been observed in some recent works \cite{Verheest09,Verheest10c} with different plasma environments. Mathematically, it is simple to prove the following property:
\begin{description}
  \item[Property:] If there exists two types of NPSWs (PPSWs) separated by a NPDL (PPDL) then there is a finite jump between the amplitudes of two types of NPSWs only when $\frac{\partial V}{\partial M}<0$ for all $M>0$ and for all $\phi<0$ ($\phi>0$).
\end{description}
For the present problem, it is easy to check that
\begin{eqnarray}
\frac{\partial V}{\partial M} = -M\bigg(\sqrt{n_{i}}-\frac{1}{\sqrt{n_{i}}}\bigg)^{2}<0
\end{eqnarray}
for all $M>0$ and for all $\phi \neq 0$. Thus all the conditions of the property are satisfied but in the positive potential side, there does not exist any jump in amplitudes between two solitary waves. More specifically, we have not found any PPDL solution which separates two types of PPSWs.

In \fref{fig:solitary and dl}(c), profiles of NPDL has been shown at $M=M_{D}$. In \fref{fig:solitary and dl}(d), profiles of NPSWs have been shown at $M=M_{D}-0.005$ and at $M=M_{D}+0.005$, respectively. The profile in \fref{fig:solitary and dl}(c) corresponding to $M=M_{D}$ is an usual double layer profile. However the solitary wave profile in \fref{fig:solitary and dl}(d)  corresponding to $M=M_{D}+0.005$ is an unusual one; its like a dais-type solitary wave profile. The jump between the amplitudes of two NPSWs separated by the NPDL is much prominent here.

In the above discussions, we have demonstrated the possible existence of solitary structures for some particular
values of the parameters of the problem without making any delimitation of the compositional parameter space for the existence of such nonlinear structures and consequently, we are unable to produce complete scenario of the present problem. So, it is desirable to construct the entire solution space or compositional parameter space showing the nature of different solitary structures present in the system. In the next section, we have considered different solution spaces of the energy integral (\ref{energy int}) with respect to $\beta_{1}$.
\section{\label{sec:solution space}Different solution spaces of the Energy integral}
\begin{figure}
\begin{center}
  \includegraphics{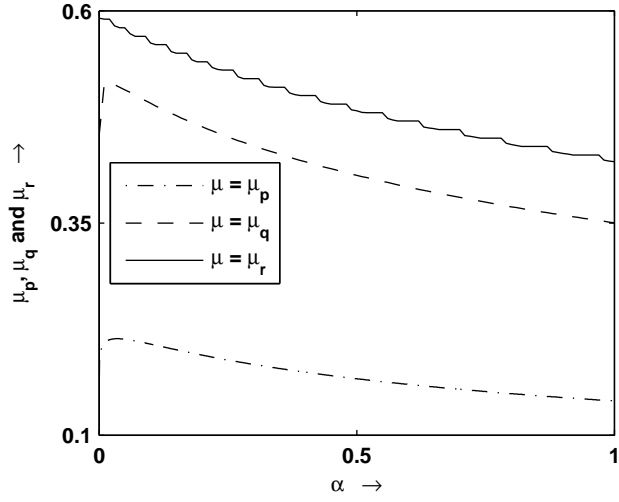}
  \caption{\label{fig:mu_pqr}$\mu_{p}$, $\mu_{q}$ and $\mu_{r}$ are plotted against $\alpha$. From this figure we see that for any value of $\alpha$ we have $\mu_{p}<\mu_{q}<\mu_{r}$.}
\end{center}
\end{figure}

\begin{figure}
\begin{center}
  \includegraphics{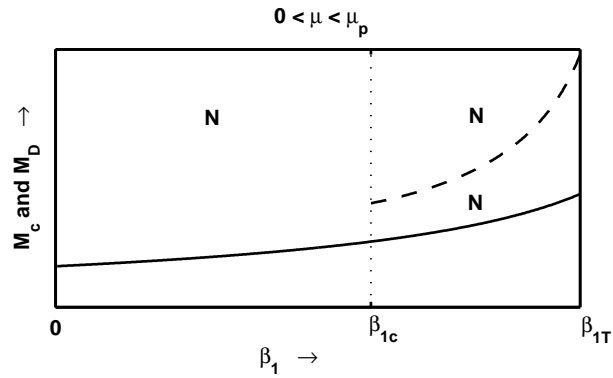}
  \caption{\label{fig:region beta_1} A graphical presentation of different solitary structures have been given with respect to two different subintervals of $\beta_{1}$ within the admissible interval of the Mach number $M$ for a particular value of $\mu$ which lies in $0 < \mu < \mu_{p}$. The curves $M = M_{c}$ (-----) and $M = M_{D} (- - -)$ are responsible for the occurrence of two subintervals of $\beta_{1}$. At any point $M=M_{D}$ one can always find a NPDL solution. This solution space has been drawn for $\alpha = 0.9$ and $\mu = 0.1$. For $\alpha = 0.9$, we have found $\mu_{p}$ lies in the neighborhood of $\mu = 0.14$ and $\beta_{1c}\approx 0.56$.}
\end{center}
\end{figure}

\begin{figure}
\begin{center}
  \includegraphics{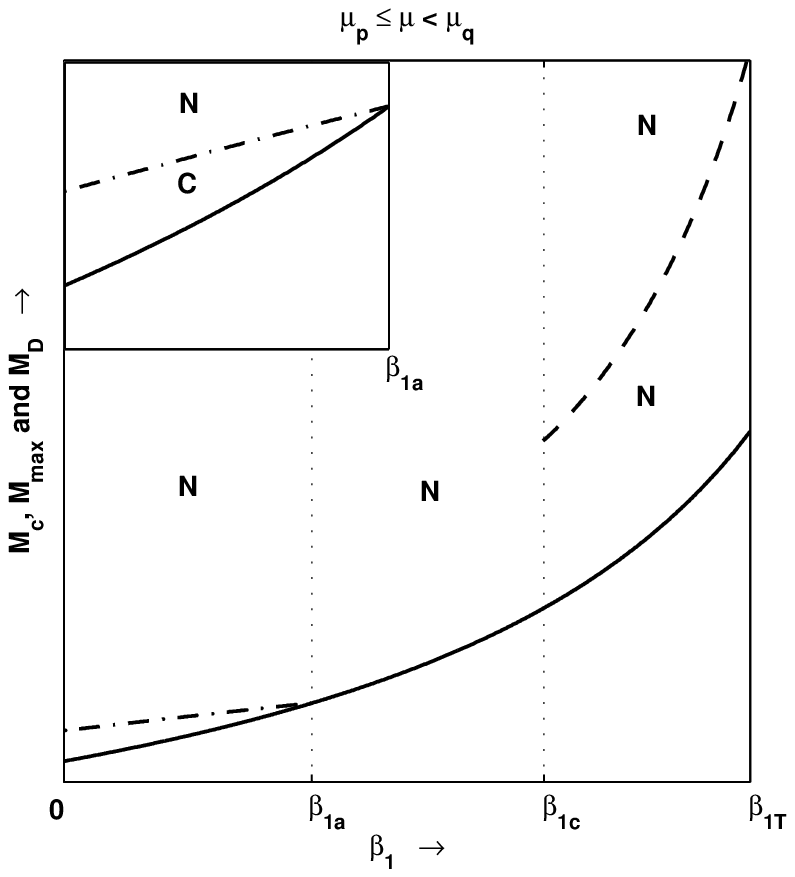}
  \caption{\label{fig:region beta_2} A graphical presentation of different solitary structures have been given with respect to three different subintervals of $\beta_{1}$ within the admissible interval of the Mach number $M$ for a particular value of $\mu$ which lies in $\mu_{p} \leq \mu < \mu_{q}$. The curves $M = M_{c}$ (-----), $M = M_{max}$ (- $\cdot$ -) and $M = M_{D} (- - -)$ are responsible for the occurrence of three subintervals of $\beta_{1}$. At any point $M=M_{D}$ one can always find a NPDL solution. The region in $0\leq \beta_{1}\leq \beta_{1a}$ has been shown in the inset. This solution space has been drawn for $\alpha = 0.9$ and $\mu = 0.25$. For $\alpha = 0.9$, we have found $\mu_{p}\approx 0.14$ and $\mu_{q}\approx 0.36$ with $\beta_{1a}\approx 0.253$ and $\beta_{1c}\approx 0.49$.}
\end{center}
\end{figure}

\begin{figure}
\begin{center}
  \includegraphics{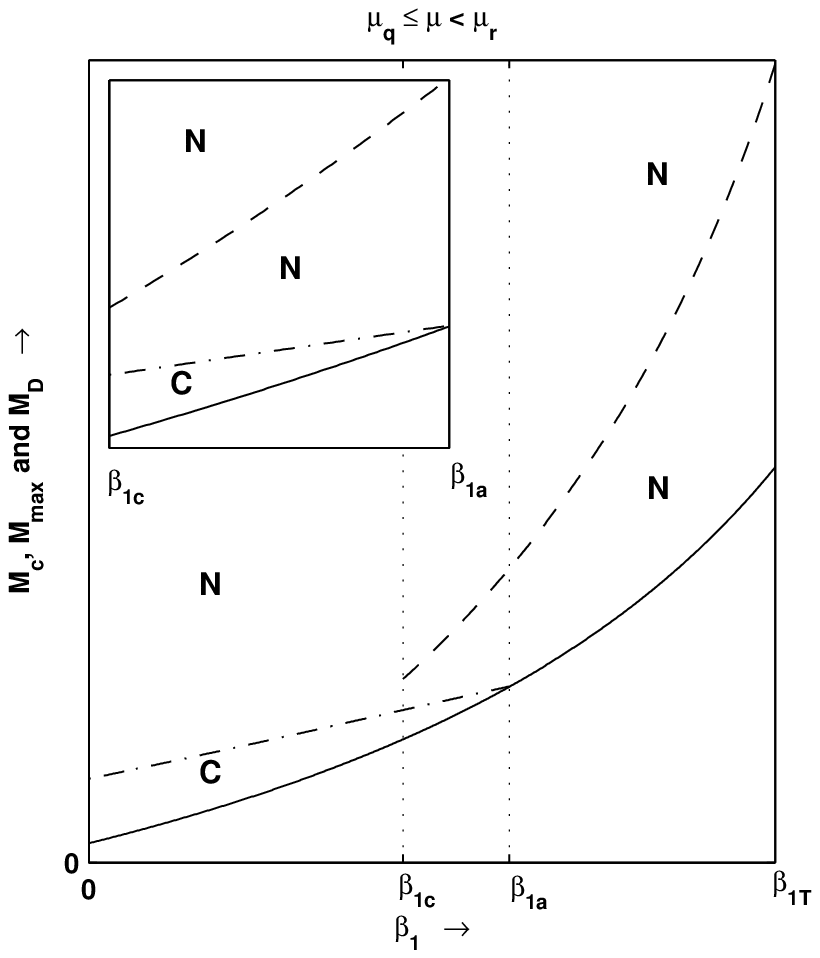}
  \caption{\label{fig:region beta_3} A graphical presentation of different solitary structures have been given with respect to four different subintervals of $\beta_{1}$ within the admissible interval of the Mach number $M$ for a particular value of $\mu$ which lies in $\mu_{q} \leq \mu < \mu_{r}$. The curves $M = M_{c}$ (-----), $M = M_{max}$ (- $\cdot$ -) and $M = M_{D} (- - -)$ are responsible for the occurrence of three subintervals of $\beta_{1}$. At any point $M=M_{D}$ one can always find a NPDL solution. This solution space has been drawn for $\alpha = 0.9$ and $\mu = 0.4$. For $\alpha = 0.9$, we have found $\mu_{q} \approx 0.36$ and $\mu_{r} \approx 0.44$ with $\beta_{1a} \approx 0.437$ and $\beta_{1c} \approx 0.352$.}
\end{center}
\end{figure}

\begin{figure}
\begin{center}
  \includegraphics{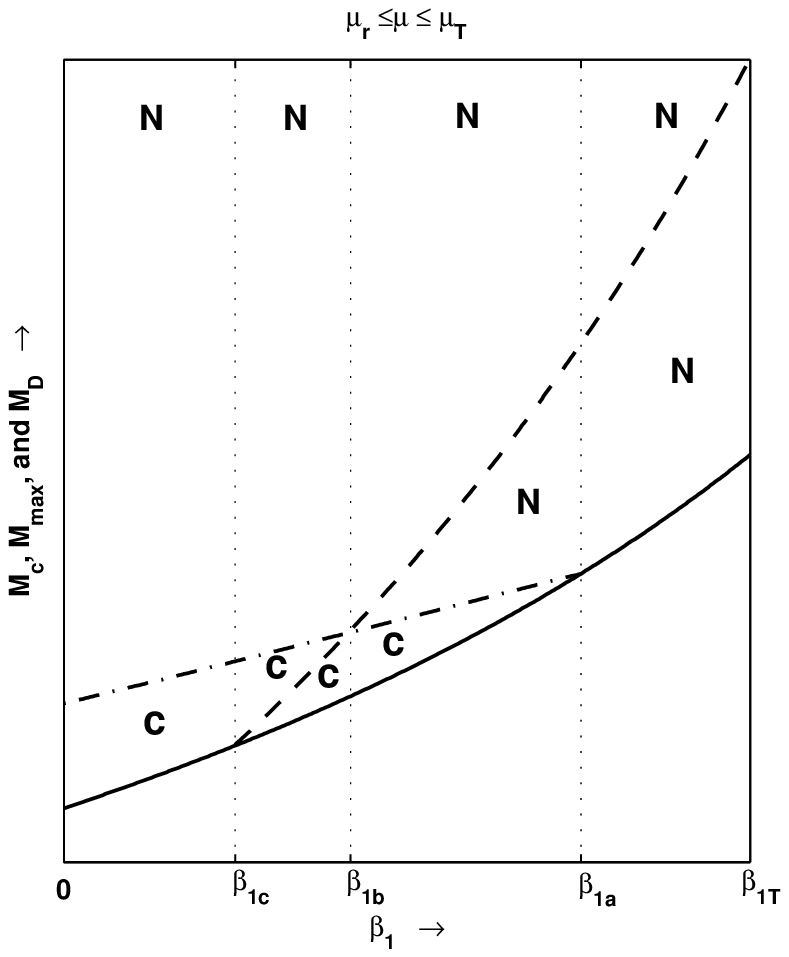}
  \caption{\label{fig:region beta_4} A graphical presentation of different solitary structures have been given with respect to four different subintervals of $\beta_{1}$ within the admissible interval of the Mach number $M$ for a particular value of $\mu$ which lies in $\mu_{r} \leq \mu \leq \mu_{T}$. The curves $M = M_{c}$ (-----), $M = M_{max}$ (- $\cdot$ -) and $M = M_{D} (- - -)$ are responsible for the occurrence of four subintervals of $\beta_{1}$. At any point $M=M_{D}$ one can always find a NPDL solution. This solution space has been drawn for $\alpha = 0.9$ and $\mu = 0.5$. For $\alpha = 0.9$, we have found $\mu_{r}$ lies in the neighborhood of $\mu = 0.44$, and $\beta_{1a} \approx 0.511$, $\beta_{1b} \approx 0.417$, $\beta_{1c} \approx 0.37$.}
\end{center}
\end{figure}

\Fref{fig:region beta_1} - \fref{fig:region beta_4} are the different compositional parameter spaces with respect to $\beta_{1}$ showing nature of solitary structures and all these figures are aimed to show the solution spaces of the energy integral (\ref{energy int}) with respect to $\beta_{1}$. To interpret \fref{fig:region beta_1} - \fref{fig:region beta_4}, we have made a general description as follows: solitary structures start to exist just above the lower curve $M = M_{c}$. For any admissible range of the parameters there always exists at least one $M>M_{c}$ such that NPSW exists thereat. $M_{max}$ is the upper bound of $M$ for the existence of PPSWs, i.e., there does not exist any PPSW if $M>M_{max}$. More explicitly, if we pick a $\beta_{1}$ and goes vertically upwards, then all intermediate $M$ bounded by $M = M_{c}$ and $M=M_{max}$ would give PPSWs. The curve $M=M_{max}$ also restrict the coexistence of both NPSWs and PPSWs, however the curve $M=M_{max}$ is unable to restrict the occurrence of all NPSWs of the present system, i.e., there exists NPSW for all $M>M_{max}$. At any point on the curve $M=M_{D}$ there exists a NPDL solution. But this NPDL solution is unable to restrict the occurrence of all NPSWs of the present system. As a result, we get two different types of NPSWs separated by the NPDL solution, in which occurrence of first type of NPSW is restricted by $M_{c}<M<M_{D}$ whereas the second type NPSW exists for all $M>M_{D}$. We have also observed a finite jump between the amplitudes of NPSWs at $M=M_{D}-\epsilon_{1}$ and at $M=M_{D}+\epsilon_{2}$, where $0<\epsilon_{1}<M_{D}-M_{c}$ and $\epsilon_{2}>0$, i.e., there is a finite jump in amplitudes of the NPSWs above and below the curve $M=M_{D}$. Now we want to define the cut off values of $\mu$ and $\beta_{1}$, which are responsible to delimit the solution space.
\begin{description}
  \item[$\mathbf{\beta_{1c}}$ : ]$\beta_{1c}$ is a cut-off value of $\beta_{1}$ such that NPDL starts to exist whenever $\beta_{1} \geq \beta_{1c}$ for any value of $\mu$ lies within the interval $0<\mu\leq \mu_{T}$, i.e., $\beta_{1} = \beta_{1c}$ is the lower bound of $\beta_{1}$ for the existence of NPDL solution. Thus, $\beta_{1c}$ is the minimum proportion of fast energetic electrons such that maximum potential difference occurs in the system and the value of $\beta_{1c}$ depends on the number of electrons residing on dust grain surface.
	\item[$\mu_{p}$ : ]$\mu_{p}$ is a cut of value of $\mu$ such that $M_{max}$ does not exist for any admissible value of $\beta_{1}$ if $\mu$ lies within the interval $0<\mu<\mu_{p}$, i.e., if $\mu \geq \mu_{p}$, there exists a value $\beta_{1}^{*}$ of $\beta_{1}$ such that $M_{max}$ exists at $\beta_{1}=\beta_{1}^{*}$, moreover, if $\beta_{1}^{*}>0$, then $M_{max}$ exists for all $\beta_{1}$ lies within the interval $0\leq \beta_{1}<\beta_{1}^{*}$. 
	\item[$\beta_{1a}$ : ]$\beta_{1a}$ is a cut-off value of $\beta_{1}$ such that $M_{max}$ exists for all $0\leq \beta_{1}\leq \beta_{1a}$ whenever $\mu \geq \mu_{p}$. Consequently, $\beta_{1} = \beta_{1a}$ is the upper bound of $\beta_{1}$ for the existence of PPSW.
\end{description}
Now, if $\beta_{1c} > \beta_{1a}$, then there exists an interval $\beta_{1a}<\beta_{1}<\beta_{1c}$ in which neither $M_{D}$ nor $M_{max}$ exist and consequently, we can define cut-off values $\mu_{q}$ and $\mu_{r}$ of $\mu$ as follows:
\begin{description}
  \item[$\mu_{q}$ : ]$\mu_{q}$ is another cut-off value of $\mu$ such that for all $\mu_{p}\leq \mu < \mu_{q}$, neither $M_{max}$ nor $M_{D}$ exist whenever $\beta_{1a}<\beta_{1}< \beta_{1c}$, i.e., for all $\mu_{p}\leq \mu < \mu_{q}$ and for all $\beta_{1a}<\beta_{1}< \beta_{1c}$ only NPSWs exist for all $M>M_{c}$.
  \item[$\mu_{r}$ : ]$\mu_{r}$ is another cut-off value of $\mu$ such that for all $\mu_{r}\leq \mu\leq \mu_{T}$, the curve $M=M_{D}$ tends to intersect the curve $M=M_{c}$ at the point $\beta_{1} = \beta_{1c}$.
\end{description}
From the definition of $\mu_{p}$, $\mu_{q}$ and $\mu_{r}$, we can numerically find the values of $\mu_{p}$, $\mu_{q}$ and $\mu_{r}$ for any value of $\alpha$. The numerical solution is shown graphically in \fref{fig:mu_pqr}. From this figure we see that for any value of $\alpha$, we can partition the entire interval of $\mu$ in the following four subintervals: (i) $0<\mu< \mu_{p}$, (ii) $\mu_{p}\leq \mu < \mu_{q}$, (iii) $\mu_{q}\leq \mu < \mu_{r}$ and (iv) $\mu_{r}\leq \mu \leq \mu_{T}$. In these subintervals of $\mu$, we have qualitatively different solution space of the energy integral (\ref{energy int}) with respect to $\beta_{1}$. The solution spaces have been shown through \fref{fig:region beta_1} - \fref{fig:region beta_4} for four different subintervals of $\mu$.

\begin{figure}
\begin{center}
  \includegraphics{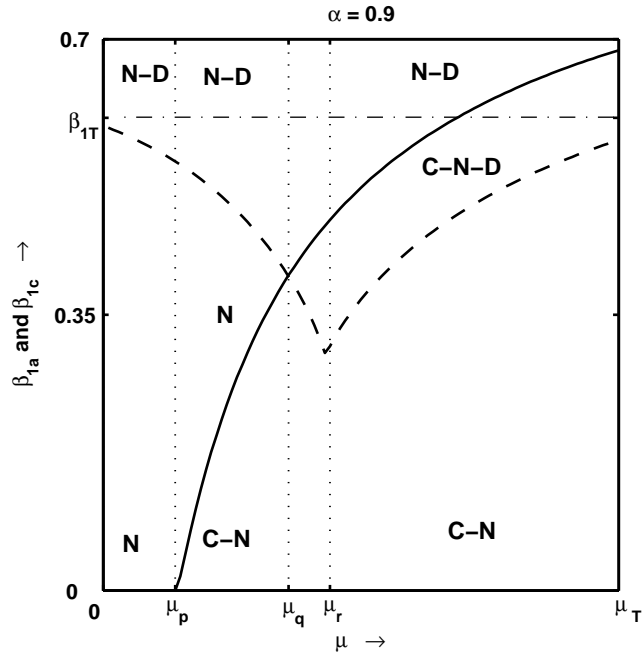}
  \caption{\label{fig:region mu_beta} Here $\beta_{1} = \beta_{1a}$ (-----) and $\beta_{1} = \beta_{1c} (- - -)$ have been plotted against $\mu$ for $\alpha = 0.9$. In this figure, actually we have shown those two curves, viz., $\beta_{1} = \beta_{1a}$ (-----) and $\beta_{1} = \beta_{1c} (- - -)$ which are responsible to divide $\mu$ into several subintervals. This figure is actually the graphical presentation of different solitary structures with respect to different subintervals of $\mu$ within the admissible interval of the nonthermal parameter $\beta_{1}$.}
\end{center}
\end{figure}

Before going to discuss the solution spaces in details, the variations of the curves $\beta_{1} = \beta_{1a}$ (-----) and $\beta_{1} = \beta_{1c} (- - -)$ have been plotted against $\mu$ in \fref{fig:region mu_beta} to demonstrate the solution spaces in true physical sense. In this figure, actually we have shown those two curves, viz., $\beta_{1} = \beta_{1a}$ and $\beta_{1} = \beta_{1c}$ which are responsible to divide $\mu$ into several subintervals. A closer look of the \fref{fig:region beta_1} - \fref{fig:region beta_4} suggests that \fref{fig:region mu_beta} effectively defined all the solution spaces as shown through \fref{fig:region beta_1} - \fref{fig:region beta_4} in more compact form provided that we have sound knowledge regarding the appropriate bounds of the Mach number for the occurrence of different types (nature) of solitary structures of the present system. Using the theory as presented in \sref{sec:mach number}, one can easily set a numerical scheme to find the appropriate bounds for the occurrence of different types (nature) of solitary structures of the present system. In \fref{fig:region mu_beta}, we have used the following terminology. C-N: region of coexistence of both PPSWs and NPSWs for $M_{c}<M\leq M_{max}$ and only NPSWs whenever $M>M_{max}$; C-N-D: region of coexistence of both PPSWs and NPSWs for $M_{c}<M\leq M_{max}$ and only NPSWs whenever $M>M_{max}$ with a NPDL at some $M=M_{D}>M_{c}$; N-D: region of existence of only NPSWs whenever $M>M_{c}$ with a NPDL at some $M=M_{D}>M_{c}$. We have found $\mu_{p}$, $\mu_{q}$ and $\mu_{r}$ lies in the neighborhood of $\mu = 0.14$, $\mu = 0.36$ and $\mu = 0.44$, respectively for $\alpha = 0.9$. From \fref{fig:region mu_beta}, we have the following observations.

For $\beta_{1}=0$, i.e., for isothermal electrons, for $0<\mu<\mu_{p}$, only NPSWs exist for all $M>M_{c}$ and coexistence of both NPSWs and PPSWs are possible for $\mu_{p}\leq \mu\leq \mu_{T}$ whenever $M_{c}<M\leq M_{max}$, whereas NPSWs exist for all $M>M_{max}$. In presence of isothermal electrons the system does not support any double layer solution. Similar facts can also be observed by considering \fref{fig:region beta_1} - \fref{fig:region beta_4} at the point $\beta_{1}=0$. So, if $\mu$ exceeds the critical value $\mu_{p}$, PPSW starts to exist for $M_{c}<M\leq M_{max}$ and attains its maximum amplitude at $M= M_{max}$ but even for increasing $\mu$ for $\mu_{p}\leq \mu\leq \mu_{T}$, the PPSW can not acquire enough strength to make a PPDL even at $M= M_{max}$. Actually, From the charge neutrality condition (\ref{charge neutrality without mu}), we see that $n_{i0}$ is a constant. Consequently, we cannot inject positive charge from outside or we cannot increase the equilibrium ion number density $n_{i0}$ and this is the reason that PPSW cannot acquire enough strength to make a PPDL even at $M= M_{max}$. On the other hand, we can increase or decrease the quantities $Z_{d}$, $n_{d0}$, $n_{e0}$ in such way that the charge neutrality condition (\ref{charge neutrality without mu}) holds good for constant equilibrium ion number density $n_{i0}$. But in any case, the amplitude of the NPSW steadily increasing for increasing Mach number $M>M_{c}$. Actually, we are unable to restrict the occurrence of NPSW for the present system, i.e., we have not found any upper bound of the Mach number which can restrict the occurrence of NPSW and this is the reason that NPSW cannot make a NPDL at any point of the compositional parameter space. However, to discuss the formation of double layer from physical point of view, we consider the following simple mathematics. Suppose $\phi_{P}$ is the amplitude of PPSW at any point of the compositional parameter space, where we have used the following terminology: if there does not exist any PPSW at some point of the compositional parameter space, then $\phi_{P}=0$. Therefore, $\phi_{P}$ is well defined as the amplitude of PPSW at any point of the compositional parameter space. Similarly, one can define $\phi_{N}$, as the amplitude of NPSW at any point of the compositional parameter space, i.e., if there does not exist any NPSW at some point of the compositional parameter space, then $\phi_{N}=0$. From simple mathematics, we get
\begin{eqnarray}
&& \phi_{P}\sim \phi_{N}=|\phi_{P}-\phi_{N}|\geq \bigg||\phi_{P}|-|\phi_{N}|\bigg|\nonumber \\
\Rightarrow && \phi_{P}\sim \phi_{N} \geq \left\{\begin{array}{c}
|\phi_{P}|-|\phi_{N}| \mbox{   if    } |\phi_{P}| \geq |\phi_{N}|\\
|\phi_{N}|-|\phi_{P}| \mbox{   if    } |\phi_{P}| \leq |\phi_{N}|
\end{array}\right.\label{phi_PN}
\end{eqnarray}
From inequality (\ref{phi_PN}), it is clear that one can get a PPDL solution at a point of the compositional parameter space if $|\phi_{P}|-|\phi_{N}|$ is maximum with $|\phi_{P}| > |\phi_{N}|$ whereas one can get a NPDL solution at a point of the compositional parameter space if $|\phi_{N}|-|\phi_{P}|$ is maximum with $|\phi_{N}| > |\phi_{P}|$. For isothermal electrons, it can be easily checked that the potential difference ($|\phi_{P}| - |\phi_{N}|$ with $|\phi_{P}| > |\phi_{N}|$) for the formation of PPDL can not attain any maximum value at any point of the compositional parameter space. Similarly, the potential difference ($|\phi_{N}| - |\phi_{P}|$ with $|\phi_{N}| > |\phi_{P}|$) for the formation of NPDL can not attain any maximum value at any point of the compositional parameter space. So, for isothermal electrons, the present system does not support any double layer solutions.

For non-zero $\beta_{1}$, the solution space as obtained in \fref{fig:region mu_beta} can be partitioned as follows: (i) $0<\mu<\mu_{p}$: For $0<\beta_{1}<\beta_{1c}$, only NPSWs are possible for all $M>M_{c}$, whereas for $\beta_{1c} \leq \beta_{1}\leq\beta_{1T}$, NPSWs are possible for all $M>M_{c}$ except $M=M_{D} (>M_{c})$. (ii) $\mu_{p}\leq \mu <\mu_{q}$: For $0<\beta_{1}\leq \beta_{1a}$, coexistence of both NPSWs and PPSWs are possible whenever $M_{c}<M\leq M_{max}$ and only NPSWs exist for all $M>M_{max}$. For $\beta_{1a}<\beta_{1}< \beta_{1c}$, only NPSWs are possible for all $M>M_{c}$. For $\beta_{1c}\leq \beta_{1}\leq \beta_{1T}$, NPSWs are possible for all $M>M_{c}$ except $M=M_{D} (>M_{c})$. (iii) $\mu_{q}\leq \mu<\mu_{r}$: For $0<\beta_{1}<\beta_{1c}$, coexistence of both NPSWs and PPSWs are possible whenever $M_{c}<M\leq M_{max}$ and only NPSWs exist for all $M>M_{max}$. For $\beta_{1c}\leq\beta_{1}\leq \beta_{1a}$, coexistence of both NPSWs and PPSWs are possible whenever $M_{c}<M\leq M_{max}$ and only NPSWs exist for all $M>M_{max}$ except the point $M=M_{D} (>M_{max})$. For $\beta_{1a}< \beta_{1}\leq \beta_{1T}$, NPSWs are possible for all $M>M_{c}$ except $M=M_{D} (>M_{c})$. (iv) $\mu_{r}\leq \mu \leq \mu_{T}$: For $0<\beta_{1}<\beta_{1c}$, coexistence of both NPSWs and PPSWs are possible whenever $M_{c}<M\leq M_{max}$ and only NPSWs exist for all $M>M_{max}$. For $\beta_{1c}\leq\beta_{1}\leq\beta_{1b}$, coexistence of both NPSWs and PPSWs are possible whenever $M_{c}<M\leq M_{max}$ and only NPSWs exist for all $M>M_{max}$ except $M=M_{D} (M_{c} < M_{D} \leq M_{max})$. For $\beta_{1b}<\beta_{1}\leq \beta_{1a}$, coexistence of both NPSWs and PPSWs are possible whenever $M_{c}<M\leq M_{max}$ and only NPSWs exist for all $M>M_{max}$ except the point $M=M_{D} (>M_{max})$. For $\beta_{1a}< \beta_{1}\leq \beta_{1T}$, NPSWs are possible for all $M>M_{c}$ except $M=M_{D} (>M_{c})$. In all these solution spaces, whenever $M_{D}>M_{c}$ or $M_{D}>M_{max}$, at the point $M=M_{D}$, one can always find a NPDL, whereas one can find the coexistence of a NPDL and a PPSW at the point $M=M_{D}$ whenever $M_{c} < M_{D} \leq M_{max}$. 

Therefore, from the above discussions, it is clear that for any physically admissible value of $\mu$, i.e., $0<\mu\leq\mu_{T}$, there exists a non-zero value of $\beta_{1}$ such that the present system supports a NPDL solution for some $M=M_{D}>M_{c}$. So, again from charge neutrality condition (\ref{charge neutrality without mu}), we see that if the density of electrons increases up to a certain value $\mu_{r}$ (see \fref{fig:region mu_beta}), minimum energetic electrons (small value of $\beta_{1}$) can produce NPDL whereas if the density of electrons tends to zero (almost depletion of electrons), more energetic electrons (higher value of $\beta_{1}$) are required to form a NPDL solution. So, we see that $\beta_{1c}$ exists for any physically admissible value of $\mu$, i.e., $0<\mu\leq\mu_{T}$ and consequently, the present system supports NPDL solution for some $M=M_{D}>M_{c}$ if $\beta_{1c}< \beta_{1} \leq \beta_{1T}$.

Again, there does not exist any $\beta_{1a}$ for $0<\mu<\mu_{p}$ and consequently, coexistence of both NPSWs and PPSWs are not possible even when nonthermal distribution of electrons becomes isothermal one, i.e., when $\beta_{1}=0$. It is also important to remember that if the value of $\beta_{1}$ increases, negative potential is stronger than positive potential and consequently, instead of getting PPSW, from inequality (\ref{phi_PN}), one can get a NPDL solution.

For $\mu_{p}\leq \mu \leq\mu_{T}$, $\beta_{1a}$ always exists and increases with increasing $\mu$. Consequently, for this interval of $\mu$, solitary structures of both polarities exist provided that $0 \leq \beta_{1} \leq \beta_{1a}$ and the region of coexistence of both NPSWs and PPSWs with respect to the nonthermal parameter increases with increasing $\mu$ lying within $\mu_{p}\leq \mu \leq\mu_{T}$. Moreover, for the values of $\mu$ lying within $\mu_{p}\leq \mu \leq \mu_{T}$, in fact, $\mu_{r}\leq \mu \leq \mu_{T}$, if $0 < \beta_{1c} \leq \beta_{1a}$, the NPSW is stronger than PPSW for $0 <\beta_{1} < \beta_{1c}$ and at $\beta_{1}=\beta_{1c}$, inequality (\ref{phi_PN}) holds good for $|\phi_{N}|>|\phi_{P}|$ and consequently, we have not only get a NPDL solution a $\beta_{1}=\beta_{1c}$ but also get a weaker PPSW at the same point of the compositional parameter space when $M=M_{D}$.    

For $\mu_{p}< \mu < \mu_{q}$, $\beta_{1c}$ is always greater than $\beta_{1a}$. Consequently, for $\beta_{1a} <\beta_{1} < \beta_{1c}$, only NPSWs exist for all $M>M_{c}$ whereas for $\mu_{q}< \mu < \mu_{T}$, $\beta_{1c}$ is always less than $\beta_{1a}$ and all types of solitary structures are possible for the present system. All the facts can also be verified by considering \fref{fig:region beta_1} - \fref{fig:region beta_4}. The physical interpretation for the formation of solitary structures in this case can be demonstrated through charge neutrality condition (\ref{charge neutrality without mu}), inequality (\ref{phi_PN}) and either considering the \fref{fig:region mu_beta} or more explicitly the figures \fref{fig:region beta_1} - \fref{fig:region beta_4}.  

Therefore, this \fref{fig:region mu_beta} is actually the graphical presentation of different solitary structures with respect to different subintervals of $\mu$ within the admissible interval of the nonthermal parameter $\beta_{1}$. 

\begin{figure}
\begin{center}
  \includegraphics{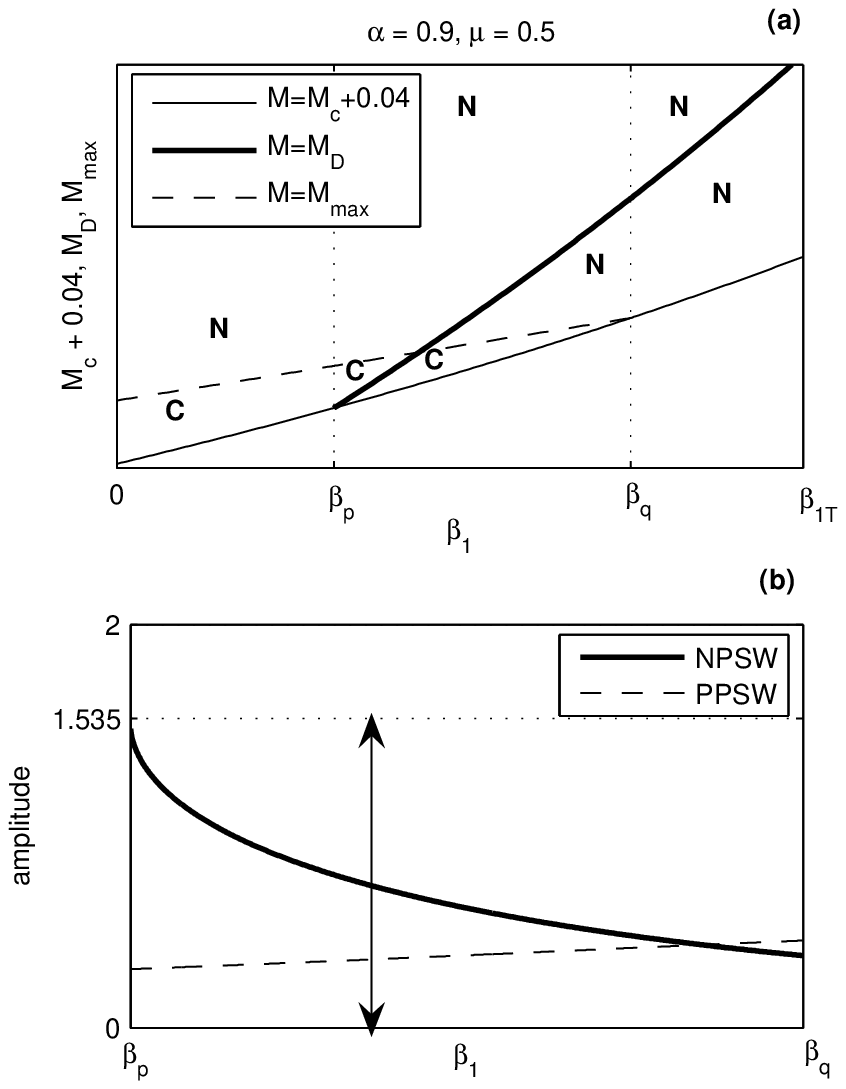}
  \caption{(a) Solution space for the present system with respect to $\beta_{1}$ has been presented, where the curve $M=M_{c}$ is omitted from the solution space as presented in \fref{fig:region beta_4}. (b)Variation in amplitude (absolute value) of both NPSW and PPSW have been shown along the curve $M=M_{c}+0.04$ for $\beta_{p}\leq \beta_{1}\leq \beta_{q}$. \label{fig:sol space amp}}
\end{center}
\end{figure}

\begin{figure}
\begin{center}
  \includegraphics{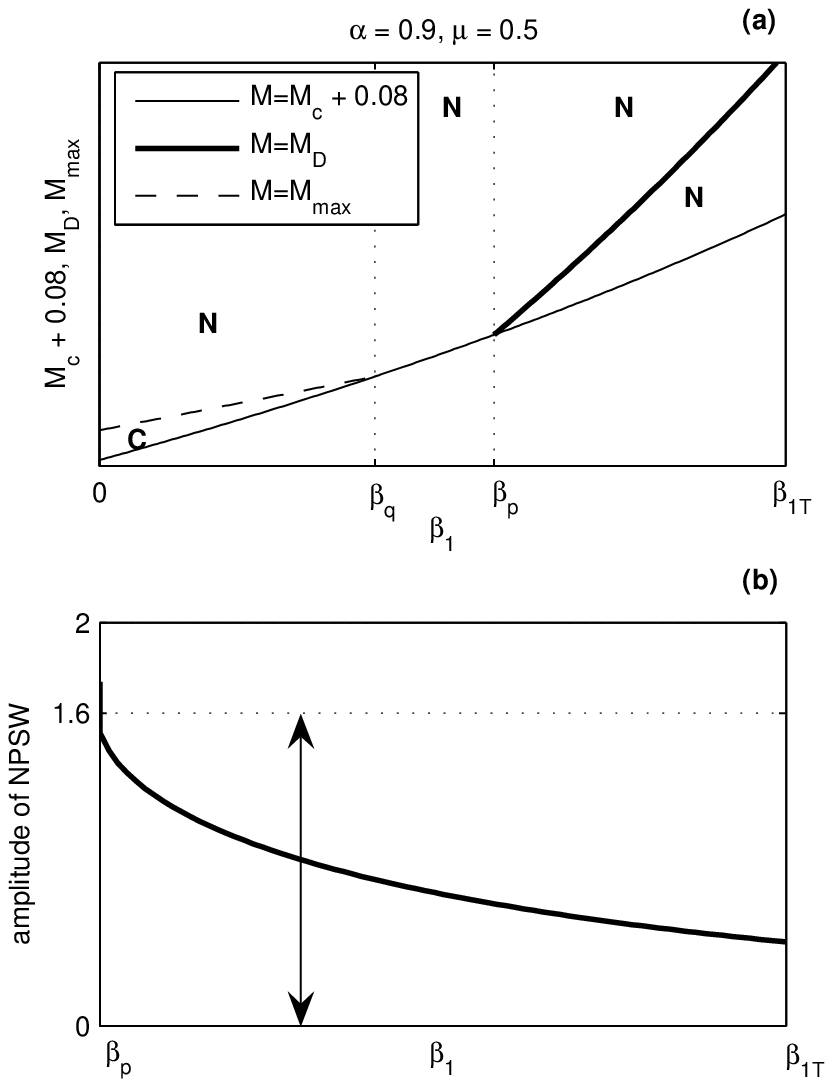}
  \caption{(a) Solution space for the present system with respect to $\beta_{1}$ has been presented, where the curve $M=M_{c}$ is omitted from the solution space as presented in \fref{fig:region beta_4}. (b)Variation in amplitude (absolute value) of NPSW have been shown along the curve $M=M_{c}+0.08$ for $\beta_{p}\leq \beta_{1}\leq \beta_{1T}$. \label{fig:sol space amp1}}
\end{center}
\end{figure}

Finally, considering any solution space, we can get new results and physical ideas for the formation of solitary structures if we move in the solution space along the family of curves parallel to the curve $M=M_{c}$. For example, we shall consider the solution space with respect to the nonthermal parameter $\beta_{1}$ for $\mu_{r}\leq \mu \leq \mu_{T}( <1)$ and if we move in the solution space along the family of curves parallel to the curve $M=M_{c}$, it is simple to understand the mathematics as well as physics for the formation of double layer solution and it is also simple to understand the relation between solitons and double layers. To be more specific, solution space for the present system with respect to $\beta_{1}$ for $\mu_{r}\leq \mu \leq \mu_{T}( <1)$ has been presented in Fig. \ref{fig:sol space amp}(a), in which the curve $M=M_{c}$ is omitted from the solution space as presented in \fref{fig:region beta_4}. Now consider the family of curves parallel to $M=M_{c}$. For instance, consider one such parallel curve for $M=M_{c}+0.04$ as shown in Fig. \ref{fig:sol space amp}(a). In this figure, $\beta_{p}$ is the value of $\beta_{1}$ where the curve $M=M_{D}$ intersects the curve $M=M_{c}+0.04$, whereas $\beta_{q}$ is the value of $\beta_{1}$ where the curve $M=M_{max}$  intersects the curve $M=M_{c}+0.04$. Fig. \ref{fig:sol space amp}(a) can be interpreted in the same way of \fref{fig:region beta_4} with $\beta_{1c}$ replaced by $\beta_{p}$ and $\beta_{1a}$ replaced by $\beta_{q}$. However, in Fig. \ref{fig:sol space amp}(a) the solitary structures exist along the curve $M=M_{c}+0.04$, specifically, (i) both NPSW and PPSW coexist for $0\leq\beta_{1}<\beta_{q}$, (ii) only NPSW exists for $\beta_{q}<\beta_{1}\leq\beta_{1T}$ and (iii) at $\beta_{1}=\beta_{p}$, a PPSW coexists with a NPDL. The variation of amplitude of those solitary waves along the curve $M=M_{c}+0.04$ for $\beta_{p}<\beta_{1}\leq\beta_{1T}$ have been shown in Fig. \ref{fig:sol space amp}(b). This figure shows that the amplitude of NPSW decrease with increasing $\beta_{1}$ for $\beta_{p}<\beta_{1}<\beta_{q}$ and the amplitudes of NPSWs are bounded by the amplitude of NPDL at $\beta_{1}=\beta_{p}$. Again, the amplitude of PPSW increases with increasing $\beta_{1}$ for $\beta_{1}>\beta_{p}$ having minimum amplitude at $\beta_{1}=\beta_{p}$. Moreover, the Fig. \ref{fig:sol space amp}(b)  shows that along the curve $M=M_{c}+0.04$, the amplitude of NPSW increases with decreasing $M$ along the curve $M=M_{c}+0.04$ for $\beta_{p}<\beta_{1}<\beta_{q}$ and ultimately, these NPSWs end with a NPDL at $\beta_{1}=\beta_{p}\approx 0.3975$. Therefore, the solitons and double layer are not two distinct nonlinear structures, i.e., double layer solution, if exists, must be the limiting structure of at least one sequence of solitons of same polarity. More specifically, existence of double layer solution implies that there must exists at least one sequence of solitary waves of same polarity having monotonically increasing amplitude converging to the double layer solution, i.e., the amplitude of the double layer solution acts as an exact upper bound or Least Upper Bound (\textit{lub}) of the amplitudes of the sequence of solitary waves of same polarity. However, we have seen in the literature that when all the parameters involved in the system assume fixed values in their respective physically admissible range, the amplitude of solitary wave increases with increasing $M$ and these solitary waves end with a double layer of same polarity, if exists. Here it is important to note that $M$ is not a function of the parameters involved in the system but is restricted by the inequality $M_{c}<M<M_{D}$, where $M=M_{D}$ corresponds to a double layer solution. So we cannot compare this case with the case of $M=M_{c}+0.04$, since $M_{c}$ is a function of the parameters involved in the system and consequently, monotonicity of $M_{c}$ entirely depends on a parameter when the other parameters assume fixed values in their respective physically admissible range. But the solitons and double layer are not two distinct nonlinear structures. Therefore, double layer solution, if exists, must be the limiting structure of at least one sequence of solitons of same polarity. In Fig. \ref{fig:sol space amp}(b), by the vertical line with both sided arrow, we mean, the amplitude of NPDL at $\beta_{1}=\beta_{p}$ for $M=M_{c}+0.04$. For $\alpha=0.9$ and $\mu=0.5$, the critical values $\beta_{p}$ and $\beta_{q}$ lies in the neighborhood of $\beta_{1}=0.3975$ and $\beta_{1}=0.4623$, respectively. Now, for the formation of NPDL solution on the curve $M=M_{c}+0.04$ at $\beta_{1}=\beta_{p}$, the negative potential (absolute value) must dominate the positive potential in the neighborhood of the point $\beta_{1}=\beta_{p}$ and the potential difference (with respect to negative potential) must be maximum thereat. From Fig. \ref{fig:sol space amp}(b), it is clear that the negative potential (absolute value) dominates the positive potential in a right neighborhood of $\beta_{1}=\beta_{p}$ and the potential difference ($|\phi_{N}|-|\phi_{P}|$ along with $|\phi_{N}| > |\phi_{P}|$) is maximum thereat. This figure shows the existence of NPDL solution at $\beta_{1}=\beta_{p}$, which is already confirmed in Fig. \ref{fig:sol space amp}(a). More specifically, from the inequality (\ref{phi_PN}), it is clear that one can get a PPDL solution at a point of the compositional parameter space if $|\phi_{P}|-|\phi_{N}|$ is maximum with $|\phi_{P}| > |\phi_{N}|$ whereas one can get a NPDL solution at a point of the compositional parameter space if $|\phi_{N}|-|\phi_{P}|$ is maximum with $|\phi_{N}| > |\phi_{P}|$. From \fref{fig:sol space amp}(b), we have found that $|\phi_{N}|-|\phi_{P}|$ is maximum with $|\phi_{N}| > |\phi_{P}|$ at $\beta_{1}=\beta_{p}$, and consequently, we can have a NPDL solution at $\beta_{1}=\beta_{p}\approx 0.3975$ for $M=M_{c}+0.04$. Next we consider the curve $M=M_{c}+0.08$ parallel to the curve $M=M_{c}$ as shown in Fig. \ref{fig:sol space amp1}(a). Here also, $\beta_{p}$ and $\beta_{q}$ are defined in the same way as in Fig. \ref{fig:sol space amp}(a). However, from Fig. \ref{fig:sol space amp1}(a), we see that there does not exist any PPSW along the curve $M=M_{c}+0.08$ for $\beta_{p}\leq \beta_{1}\leq\beta_{1T}$. So according to the terminology $\phi_{P}=0$ along the curve $M=M_{c}+0.08$ for $\beta_{p}\leq \beta_{1}\leq\beta_{1T}$. Consequently, from inequality (\ref{phi_PN}), we have NPDL solution at $\beta_{1}=\beta_{p}\approx 0.4234$ for $M=M_{c}+0.08$. This fact is clear from \fref{fig:sol space amp1}(b).

\section{\label{sec:conclusion}Summary and Discussions}
In the present paper, we have investigated DIASWs and DIADLs in a dusty plasma system consisting of adiabatic ions, nonthermal electrons and negatively charged dust grains. Investigations have been made by going through the entire solution space of the energy integral by considering the entire range of the parameters involved in the system. Our aim is to delimit the parameter $\beta_{1}$ depending on the nature of existence of DIASWs and DIADLs.

Therefore, for any physically admissible values of the parameters of the system, specifically, for any value of $\mu$ and any value of $\beta_{1}$, NPSW exists for all $M>M_{c}$ except $M=M_{D}$, where $M_{c}$ is the lower bound of Mach number $M$, i.e., solitary wave and/or double layer solutions of the energy integral start to exist for $M>M_{c}$ and $M_{D}$ is the Mach number corresponding to a NPDL solution. However, if the parameter $\mu$ exceeds a critical value $\mu_{p}$, PPSWs exist for all $\mu_{p} \leq \mu < 1$ whenever the Mach number lies within the interval $M_{c} < M \leq M_{max}$, where $M_{max}$ is the upper bound of $M$ which is well-defined only when the system supports PPSWs. Therefore, the coexistence of both PPSWs and NPSWs is possible for all $\mu_{p} \leq \mu <1$ whenever $M_{c} < M \leq M_{max}$, but NPSWs still exist for $M>M_{max}$.

For nonthermal electrons, NPDL starts to occur whenever the nonthermal parameter exceeds a critical value. However this double layer solution is unable to restrict the occurrence of NPSWs. As a result, two different types of NPSWs have been observed, in which occurrence of first type of NPSW is restricted by $M_{c}<M<M_{D}$ whereas the second type NPSW exists for all $M>M_{D}$, where $M_{D}$ is the Mach number corresponding to a NPDL. A finite jump between the amplitudes of NPSWs at $M=M_{D}-\epsilon$ and at $M=M_{D}+\epsilon$ has been observed, where $\epsilon$ is a sufficiently small positive quantity. The amplitude of NPSW for $M>M_{D}$ is much greater than the amplitude of the NPDL solution at $M=M_{D}$ as well as the amplitude of NPSW for $M<M_{D}$, i.e., there is a jump in amplitudes of the NPSWs above and below the curve $M=M_{D}$. However, there is no jump in amplitudes of NPSWs above and below the curve $M=M_{max}$.

In most of the earlier works, dust ion acoustic solitary structures have been investigated with the help of Maxwellian velocity distribution function for electrons. However, the dusty plasma with nonthermally/suprathermally  distributed electrons observed in a number of heliospheric environments \cite{Verheest00,Shukla02,Asbridge68,Feldman83,Lundin89,Futaana03}. Therefore, the present paper gives the complete scenario of dust ion acoustic solitary structures in a dusty plasma system in which lighter species (electrons) is nonthermally distributed. This paper is also helpful for understanding the formation of dust ion acoustic solitary structures from different physical and mathematical aspects.
\ack
One of the authors (Animesh Das) is thankful to State Government Departmental Fellowship Scheme for providing research support.

\newpage


\end{document}